\documentclass[aps,prb,twocolumn,superscriptaddress,letter]{revtex4}
\usepackage{amsfonts}
\usepackage{amsmath}
\usepackage{color}
\usepackage{graphicx}
\usepackage{dcolumn}
\usepackage{bm}

\begin{document}

\title{Quantum spin liquid near Mott transition with fermionized $\pi $%
-vortices}
\author{Su-Peng Kou}
\thanks{Corresponding author}
\email{spkou@bnu.edu.cn}
\affiliation{Department of Physics, Beijing Normal University, Beijing 100875, China}
\author{Lan-Feng Liu }
\affiliation{Department of Physics, Beijing Normal University, Beijing 100875, China}
\author{Jing He}
\affiliation{Department of Physics, Beijing Normal University, Beijing 100875, China}
\author{Ya-Jie Wu}
\affiliation{Department of Physics, Beijing Normal University, Beijing 100875, China}

\begin{abstract}
In this paper, we study the non-magnetic insulator state near Mott
transition of 2D $\pi $-flux Hubbard model on square lattice and find that
such non-magnetic insulator state is quantum spin liquid state with nodal
fermionic excitations - nodal spin liquid (NSL). When there exists small
easy-plane anisotropic energy, the ground state becomes $Z_{2}$ topological
spin liquid (TSL) with full gapped excitations. The $U(1)\times U(1)$
mutual--Chern-Simons (MCS) theory is obtained to describe the low energy
physics of NSL and TSL.

\end{abstract}

\maketitle

\section{Introduction}

People have been looking for quantum spin liquid states in frustrated spin
models for more than 20 years \cite{Fazekas,wen}. For example, various
approaches show that quantum spin liquids may exist in two-dimensional (2D) $%
S=1/2$ \textrm{J}$_{1}$\textrm{-J}$_{2}$ model or the Heisenberg model on
Kagom\'{e} lattice\cite{k1,k2,k3,k4,k5}. In these models, the quantum spin
liquids are accessed (in principle) by appropriate frustrating interactions.
However, the nature of the quantum disordered ground state is still much
debated\cite{russi,Dagotto,Sano,Schultz,Sorella,cap,j1j2,zhang}. People have
guessed that the quantum disordered state may be either algebra spin liquid
state\cite{wen1,he}, or Z2 spin liquid state\cite{wen2} or chiral spin liquid%
\cite{wen3,wen4}.

On the other hand, the experiments in the organic material $\mathrm{\kappa
-(BEDT-TTF)}_{2}\mathrm{CU}_{2}\mathrm{(CN)}_{3}$ indicate the realization
of the quantum spin liquid states\cite{Shimizu,Kawamoto,Kurosaki}. Then, $%
U(1)$ and $SU(2)$ slave-roton theories of the Hubbard model were formulated
on the triangular or honeycomb lattices\cite{Lee,Hermele}. Recently, the
quantum spin liquid state near Mott transition of the Hubbard model on
honeycomb lattice has been conformed by different approaches\cite%
{meng,kou4,z21,z22,z23,z24,z25}. In particular, in Ref.\cite{meng}, quantum
spin liquid state has been predicted by quantum Monte Carlo (QMC)
simulation.
\begin{figure}[tbp]
\includegraphics[width=0.3\textwidth]{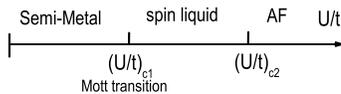}
\caption{The scheme of the spin liquid state near Mott transition of the $%
\protect\pi $-flux Hubbard model }
\end{figure}

In Ref.\cite{kou1}, the possible non-magnetic state in the nodal
antiferromagnetic (AF) insulator (an insulator with AF spin density wave
ordering and massive Dirac fermionic excitations) is predicted to near the
Mott transition of the $\pi $-flux Hubbard model (See Fig.1). Such type of
quantum disordered states in bipartite lattices is also not driven by
frustrations, as people have done in varied spin models. Instead, they come
from quantum fluctuations of a relatively small effective spin-moments near
Mott transition. What's the nature of the possible non-magnetic state here?
In this paper we will answer the question. We find that for the isotropic
case,\textbf{\ }the non-magnetic state with \textrm{SU(2) }spin rotation
symmetry is a new type of spin liquid - \emph{nodal spin liquid (NSL) with
gapless fermionic excitations and roton-like excitations; }on the other
hand, for the anisotropic case by adding small easy-plane anisotropic term,
the non-magnetic state with weakly breaking \textrm{SU(2) }spin rotation
symmetry becomes a $Z_{2}$ topological spin liquid (TSL) with topological
degenerate ground states.\emph{\ }

The paper is organized as follows. In Sec.II, we study the spin fluctuations
of the nodal AF insulator (NAI) in $\pi $-flux Hubbard model on square
lattice based on a formulation by keeping spin rotation symmetry. In
Sec.III, we study the properties of half skyrmions - topological solitons of
the nodal AF insulator and show their induced quantum numbers and
statistics. In Sev.IV, nodal spin liquid is proposed to be the ground state
of the quantum non-magnetic insulator state in the nodal AF insulator. In
this section, we use $U(1)\times U(1)$ mutual--Chern-Simons theory to learn
the properties of NSL and TSL. Finally, the conclusions are given in Sec.V.

\section{Non-magnetic state in nodal AF insulator of the Hubbard model on $%
\protect\pi $-flux lattice}

In this paper we will focus on the $\pi $-flux Hubbard model on square
lattice\cite{flux}. The Hamiltonian of it is
\begin{equation}
\mathcal{H}=-\sum\limits_{\langle i,j\rangle }\left( t_{ij}\hat{c}%
_{i}^{\dagger }\hat{c}_{j}+h.c.\right) +U\sum_{i}\hat{n}_{i\uparrow }\hat{n}%
_{i\downarrow }.
\end{equation}%
Here $\hat{c}_{i}=(\hat{c}_{i\uparrow },\hat{c}_{i\downarrow })^{T}$ are
defined as electronic annihilation operators. $U$ is the on-site Coulomb
repulsion. $\langle i,j\rangle $ denotes two sites on a nearest-neighbor
link. $\hat{n}_{i\uparrow }$ and $\hat{n}_{i\downarrow }$ are the number
operators of electrons at site $i$ with up-spin and down-spin, respectively.
There is a $\pi $-flux phase when a particle hops around a plaquette in a $%
\pi $-flux lattice. So the nearest-neighbor hopping $t_{i,j}$ in a $\pi $%
-flux lattice could be chosen as $t_{i,i+\hat{x}}=t,t_{i,i+\hat{y}}=te^{\pm i%
\frac{\pi }{2}}$\cite{hsu}. Although $\pi $-flux phase does not break
translational symmetry, we may still divide the square lattice into two
sublattices, $A$ and $B$. After transforming the hopping term into momentum
space, we obtain $\xi _{\mathbf{k}}=\pm \sqrt{4t^{2}\left( \cos
^{2}k_{x}+\cos ^{2}k_{y}\right) }.$ So there exist two nodal fermi-points at
$\mathbf{k}_{1}=(\frac{\pi }{2},\frac{\pi }{2}),$ $\mathbf{k}_{2}=(\frac{\pi
}{2},-\frac{\pi }{2})$ and the spectrum of fermions becomes linear in the
vicinity of the two nodal points. In the non-interacting limit, the $\pi $%
-flux Hubbard model is reduced into a free fermion model with nodal fermions
at half filling.

Because the Hubbard model on bipartite lattices is unstable against
antiferromagnetic instability, at half-filling, the ground state may be an
insulator with AF SDW\ order with increasing interacting strength. Such AF
SDW order is described by the following mean field order parameter $\langle
(-1)^{i}\hat{c}_{i}^{\dag }\sigma ^{z}\hat{c}_{i}\rangle =\frac{M}{2}.$ Here
$M$ is the staggered magnetization. Then in the mean field theory, the
Hamiltonian of the 2D $\pi $-flux Hubbard model is obtained as%
\begin{equation}
\mathcal{H}=-\sum\limits_{\left\langle ij\right\rangle }(t_{i,j}\hat{c}%
_{i}^{\dagger }\hat{c}_{j}+h.c.)-\sum_{i}\left( -1\right) ^{i}\Delta \hat{c}%
_{i}^{\dagger }\mathbf{\sigma }_{z}\hat{c}_{i}  \label{m}
\end{equation}%
where $\Delta =\frac{UM}{2}$ leads to the energy gap of electrons and $%
\sigma _{z}$ is the Pauli matrix. After diagonalization, the spectrum of the
electrons is obtained as $E_{\mathbf{k}}=\pm \sqrt{\xi _{\mathbf{k}%
}^{2}+\Delta ^{2}}.$ By minimizing the free energy at zero temperature, the
self-consistent equation of (\ref{m}) is reduced into $\frac{1}{N}%
\sum\limits_{\mathbf{k}}\frac{U}{2E_{\mathbf{k}}}=1$ where $N$ is the number
of the particles.

From mean field approach, one may see that the MI transition of the $\pi $%
-flux Hubbard model occurs at a critical value about $U/t\simeq 3.11$. In
the weakly coupling limit $\left( U/t<3.11\right) $, the ground state is a
semi-metal (SM) with nodal fermi-points. In the strong coupling region $%
\left( U/t>3.11\right) $, due to $M\neq 0$, the ground state becomes an
insulator with relativistic massive fermionic excitations. However, the
non-zero value of $M$ only means the existence of effective spin moments. It
does not necessarily imply that the ground state of NAI is a long range AF
order in the mean field theory. Thus one needs to examine stability of
magnetic order against quantum fluctuations of effective spin moments based
on a formulation by keeping spin rotation symmetry, $\sigma _{z}\rightarrow
\mathbf{\Omega }\cdot \mathbf{\sigma .}$

Within the haldane's mapping, the spins are parametrized as $\mathbf{\Omega }%
_{i}=(-1)^{i}\mathbf{n}_{i}\sqrt{1-\mathbf{L}_{i}^{2}}+\mathbf{L}_{i}$\cite%
{Haldane,Auerbach,dup,dupuis}. Here $\mathbf{n}_{i}$ is the AF order
parameter and $\left\vert \mathbf{n}_{i}\right\vert =1$, $\mathbf{L}_{i}$ is
the transverse canting field, which is chosen to $\mathbf{L}_{i}\cdot
\mathbf{n}_{i}=0.$ By replacing the electronic operators $\hat{c}%
_{i}^{\dagger }$ and $\hat{c}_{j}$ by Grassmann variables $c_{i}^{\ast }$
and $c_{j}$, we get the effective Lagrangian with spin rotation symmetry as%
\begin{equation}
\mathcal{L}_{\mathrm{eff}}=\sum_{i}c_{i}^{\ast }\partial _{\tau
}c_{i}-\sum\limits_{\left\langle ij\right\rangle }(t_{i,j}c_{i}^{\ast
}c_{j}+h.c.)-\sum_{i}\left( -1\right) ^{i}\Delta c_{i}^{\ast }\mathbf{\Omega
}_{i}\mathbf{\cdot \sigma }c_{i}.
\end{equation}

After integrating the massive fermions and transverse canting field,\ an
effective $\mathrm{O(3)}$\ nonlinear $\sigma $\ model appears that describes
the long-wavelength spin fluctuations with spin rotation symmetry\cite%
{wen,dup,dupuis,Schulz,weng,kou1,theta}%
\begin{equation}
\mathcal{L}_{s}=\frac{1}{2g}\left( \partial _{\mu }\mathbf{n}\right) ^{2}.
\label{cons}
\end{equation}%
The coupling constant is obtained as $g=\sqrt{\frac{1}{\chi ^{\perp }\rho
_{s}}}$ where
\begin{widetext}
\begin{equation}
\rho _{s}=\frac{1}{N}\sum \limits_{\mathbf{k}}\frac{t^{2}[\cos \left(
2k_{x}\right) \left( \Delta ^{2}+8t^{2}+4t^{2}\cos \left( 2k_{y}\right)
\right) +\Delta ^{2}+3t^{2}+t^{2}\cos \left( 4k_{x}\right) ]}{2(\left \vert
\xi _{\mathbf{k}}\right \vert ^{2}+\Delta ^{2})^{\frac{3}{2}}}
\end{equation}%
\end{widetext}and the transverse spin susceptibility is
\begin{equation}
\chi ^{\perp }=[(\frac{1}{N}\sum\limits_{\mathbf{k}}\frac{\Delta ^{2}}{%
4(\left\vert \xi _{\mathbf{k}}\right\vert ^{2}+\Delta ^{2})^{\frac{3}{2}}}%
)^{-1}-2U]^{-1}.
\end{equation}%
See detailed calculations in appendix. In the paper we will set the spin
velocity to be unit.

The coupling constant $g$ of the $\pi $-flux Hubbard model has been obtained
in Ref.\cite{kou1}. In Ref.\cite{kou1}, a quantum critical point at $g_{c}=%
\frac{4\pi }{\Lambda }$ was found that corresponds to $U/t\simeq 4.26$ of
the $\pi $-flux Hubbard model. For the case of $g>g_{c},$ a quantum
non-magnetic insulator (NMI) state with gapped spin excitations as $%
m_{s}=4\pi (\frac{1}{g_{c}}-\frac{1}{g})$ appears that corresponds to the
region of $3.11<U/t<4.26;$ For the case of $g<g_{c},$ the ground state is a
long range AF order that corresponds to the region of $U/t>4.26$. Here the
momentum cutoff is introduced as $\Lambda =\min (a^{-1},$ $2\Delta )$ \cite%
{dup,dupuis} ($a$ is lattice constant). The existence of a non-magnetic
insulator state provides an alternative candidate for finding spin liquid
state. An interesting issue is the nature of the non-magnetic insulator. Is
it a valence-band crystal, or algebra spin liquid state, or a new type of
quantum state? In the following parts we will answer the question.

\section{Half-skyrmion as fermionic excitation}

To learn the nature of the non-magnetic insulator, one may study its
excitations. In the non-magnetic insulator, we get the effective model of
massive spin-1 excitations
\begin{equation}
\mathcal{L}_{\mathbf{s}}=\frac{1}{2g}\left[ (\partial _{\mu }\mathbf{n}%
)^{2}+m_{s}^{2}\mathbf{n}^{2}\right] .
\end{equation}%
Or using the CP(1) representation, we have
\begin{equation}
\mathcal{L}_{\mathbf{s}}=\frac{2}{g}\left[ |(\partial _{\mu }-ia_{\mu })%
\mathbf{z}|^{2}+m_{z}^{2}\mathbf{z}^{2}\right]
\end{equation}%
where $\mathbf{z}$ is a bosonic spinon, $\mathbf{z}=\left(
z_{1},z_{2}\right) ,$ $\mathbf{n}_{i}=\mathbf{\bar{z}}_{i}\mathbf{\sigma z}%
_{i}\mathbf{,}$ $\mathbf{\bar{z}z=1,}$ $a_{\mu }\equiv -\frac{i}{2}(\mathbf{%
\bar{z}}\partial _{\mu }\mathbf{z}-\partial _{\mu }\mathbf{\bar{z}z})$. Here
$a_{\mu }$\ is introduced as an assistant gauge field. Because the bosonic
spinons have mass gap, we may integrate them and get $\mathcal{L}_{\mathrm{%
eff}}=\frac{1}{4e_{a}^{2}}(\partial _{\mu }a_{\nu })^{2}$ with $%
e_{a}^{2}\sim m_{z}=\frac{1}{2}m_{s}$. Due to the instanton effect, the
gauge field $a_{\mu }$ obtains a mass gap and bosonic spinons that couple
the gauge field $a_{\mu }$ are confined. Now the lowest energy excitations
are gapped spin wave. However, the answer is not quite right. Applying the
Oshikawa's commensurability condition and Hastings' theorem to the present
case, the quantum state without spontaneously symmetry breaking can be
either a topological order with degenerate ground state or a uniform ground
state with triplet excitation gap must be accompanied with other gapless
excitations\cite{Oshikawa,hua}. In this paper we indeed find that
non-magnetic insulator is either topological spin liquid with degenerate
ground state or a nodal spin liquid with gapless fermionic excitations.

\subsection{half-skyrmion}

In the following parts, we focus on the half-skyrmions (topological vortices
with half topological charge),
\begin{equation}
\mathcal{Q}=\int \mathbf{d^{2}\mathbf{r}}\frac{1}{4\pi }\epsilon _{0\nu
\lambda }\mathbf{n\cdot \partial }^{\nu }\mathbf{n\times \partial }^{\lambda
}\mathbf{n=\pm }\frac{1}{2}.
\end{equation}%
In the vector $\mathbf{n}$ representation, the solutions of the
half-skyrmion of the continuum limit are \cite%
{pol,meron,half,ng,ots,weng1,kou2,kou}%
\begin{eqnarray}
\mathbf{n}_{\mathrm{hs}} &=&(\frac{\lambda (x-x_{0})}{\sqrt{|\mathbf{r}-%
\mathbf{r}_{0}|^{2}+\lambda ^{2}}},\text{ }\pm \frac{\lambda (y-y_{0})}{%
\sqrt{|\mathbf{r}-\mathbf{r}_{0}|^{2}+\lambda ^{2}}},\text{ } \\
&&\pm \frac{\lambda }{\sqrt{|\mathbf{r}-\mathbf{r}_{0}|^{2}+\lambda ^{2}}}).
\notag
\end{eqnarray}%
Here $\lambda $ is the radius of the half-skyrmion at $\mathbf{r}%
_{0}=(x_{0},y_{0})$. Inside the core $|\mathbf{r}-\mathbf{r}%
_{0}|^{2}<\lambda ,$ the spin is polarized; outside it $|\mathbf{r}-\mathbf{r%
}_{0}|^{2}>\lambda $, one gets a vortex-like spin configuration on $XY$
plane. To stabilize topological vortices (half-skyrmions), a small
easy-plane anisotropic term should be added to the original model
phenomenally, $H^{\prime }=\kappa \sum_{i}\mathbf{n}_{i}^{2}$ ($\kappa >0,$ $%
\frac{\kappa }{t}\ll 1$).

From the exact solutions, there exist two types of merons : one has a
up-spin polarized core $\mathcal{Q}=-\frac{1}{2},$%
\begin{eqnarray}
\left( \mathbf{n}_{\mathrm{hs}}\right) _{1} &=&(\frac{\lambda (x-x_{0})}{%
\sqrt{|\mathbf{r}-\mathbf{r}_{0}|^{2}+\lambda ^{2}}},\text{ }\frac{\lambda
(y-y_{0})}{\sqrt{|\mathbf{r}-\mathbf{r}_{0}|^{2}+\lambda ^{2}}}, \\
&&\text{ }\frac{\lambda }{\sqrt{|\mathbf{r}-\mathbf{r}_{0}|^{2}+\lambda ^{2}}%
});  \notag
\end{eqnarray}%
the other a down-spin polarized core $\mathcal{Q}=\frac{1}{2}$,%
\begin{eqnarray}
\left( \mathbf{n}_{\mathrm{hs}}\right) _{2} &=&(\frac{\lambda (x-x_{0})}{%
\sqrt{|\mathbf{r}-\mathbf{r}_{0}|^{2}+\lambda ^{2}}},\text{ }\frac{\lambda
(y-y_{0})}{\sqrt{|\mathbf{r}-\mathbf{r}_{0}|^{2}+\lambda ^{2}}}, \\
&&\text{ }-\frac{\lambda }{\sqrt{|\mathbf{r}-\mathbf{r}_{0}|^{2}+\lambda ^{2}%
}}).  \notag
\end{eqnarray}%
Fig.2 and Fig.3 show the two types of merons. From them, one can see that
the topological charge of a half-skyrmion is determined by\emph{\ both the
spin configuration and the polarized direction of AF order in the core}.

\begin{figure}[tbph]
\includegraphics[width = 10cm]{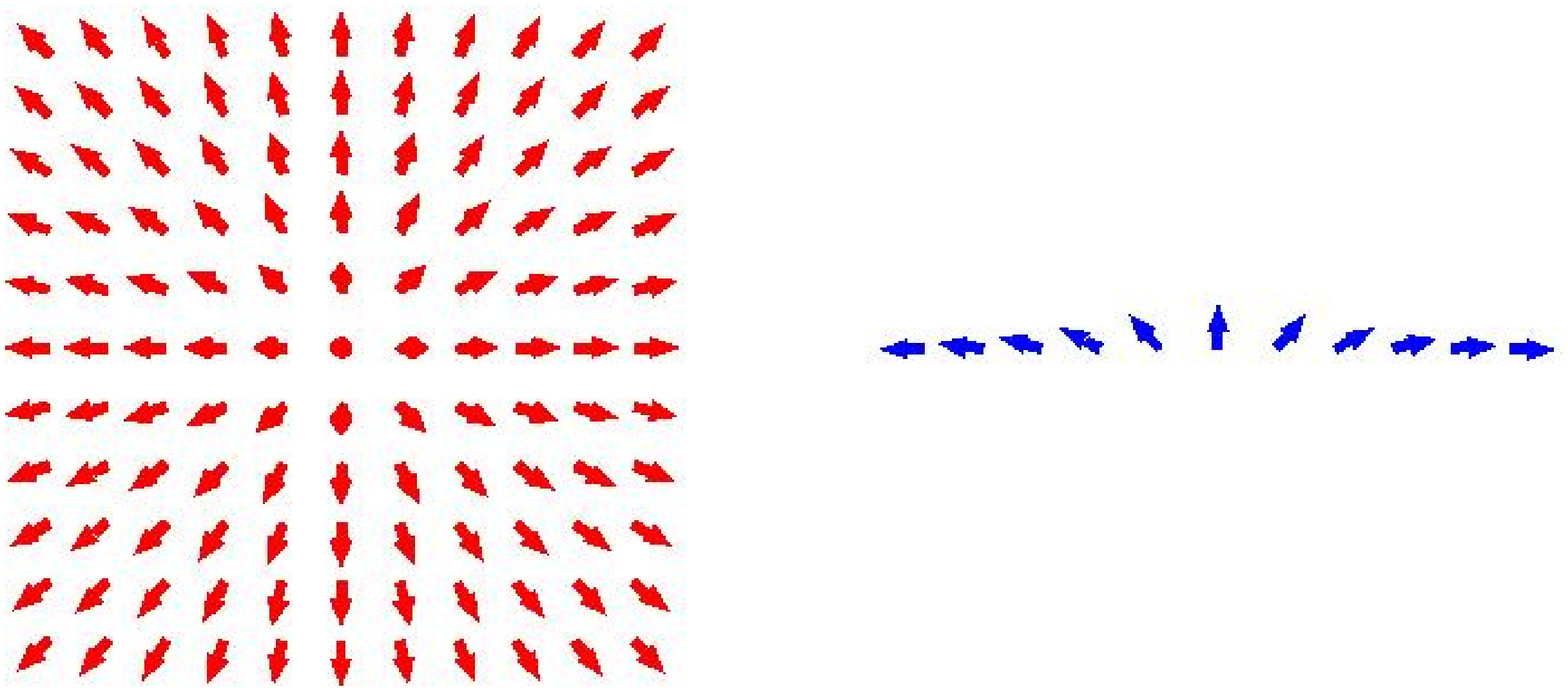}
\caption{The scheme of $\left( \mathbf{n}_{\mathrm{hs}}\right) _{1}$, the
meron with topological charge $-1/2$. up-spins locate at the center.}
\end{figure}

\begin{figure}[tbph]
\includegraphics[width =
10cm]{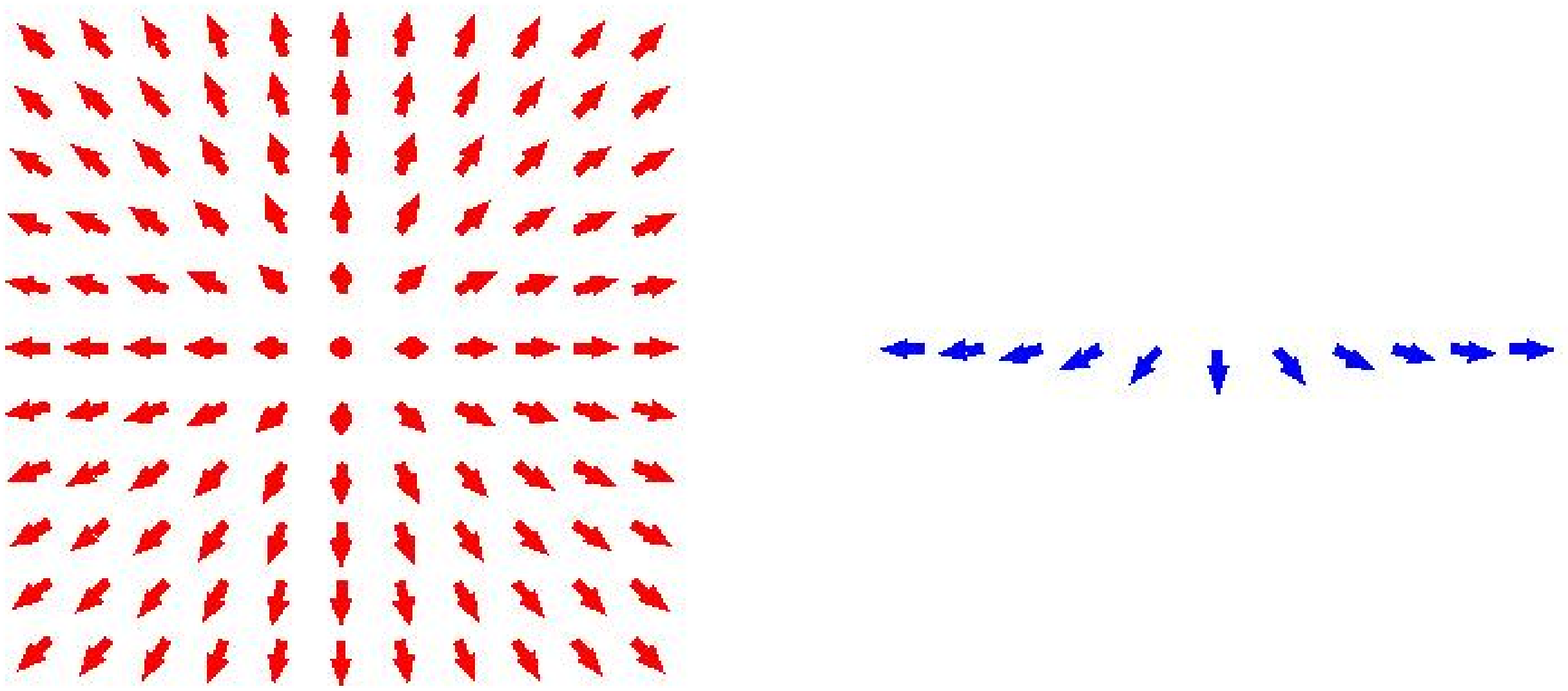} \caption{The scheme of
$\left(\mathbf{n}_{\mathrm{hs}}\right) _{2}$, the meron with
topological charge $1/2$. Down-spins locate at the center.}
\end{figure}

Also, there are two types of anti-merons : one has a up-spin polarized core $%
\mathcal{Q}=\frac{1}{2}$,%
\begin{eqnarray}
\left( \mathbf{n}_{\mathrm{hs}}\right) _{3} &=&(\frac{\lambda (x-x_{0})}{%
\sqrt{|\mathbf{r}-\mathbf{r}_{0}|^{2}+\lambda ^{2}}},\text{ }-\frac{\lambda
(y-y_{0})}{\sqrt{|\mathbf{r}-\mathbf{r}_{0}|^{2}+\lambda ^{2}}},\text{ } \\
&&\frac{\lambda }{\sqrt{|\mathbf{r}-\mathbf{r}_{0}|^{2}+\lambda ^{2}}});
\notag
\end{eqnarray}%
the other has a down-spin polarized core $\mathcal{Q}=-\frac{1}{2}$,%
\begin{eqnarray}
\left( \mathbf{n}_{\mathrm{hs}}\right) _{4} &=&(\frac{\lambda (x-x_{0})}{%
\sqrt{|\mathbf{r}-\mathbf{r}_{0}|^{2}+\lambda ^{2}}},\text{ }-\frac{\lambda
(y-y_{0})}{\sqrt{|\mathbf{r}-\mathbf{r}_{0}|^{2}+\lambda ^{2}}}, \\
&&\text{ }-\frac{\lambda }{\sqrt{|\mathbf{r}-\mathbf{r}_{0}|^{2}+\lambda ^{2}%
}}).  \notag
\end{eqnarray}

In AF ordered state, the mass of half-skyrmion $m_{hs}$ is associated with
the ordered staggered moment :
\begin{equation}
m_{hs}=2\pi \tilde{\rho}_{s}+E_{\mathrm{core}}\text{ (}g<g_{c}\text{)}
\end{equation}%
where $\tilde{\rho}_{s}=\rho _{s}(1-\frac{g}{g_{c}})$ is the renormalized
spin stiffness and $E_{\mathrm{core}}$ is the core energy. For the isotropic
case $\kappa =0,$ the energy of the half-skyrmion does not depend on its
radius $\lambda $ and the core energy is zero, $E_{\mathrm{core}}=0$. For
the anisotropy case $\kappa >0,$ the scale invariance is broken at $%
H^{\prime }=\kappa \sum_{i}\mathbf{n}_{i}^{2}\sim H_{\mathrm{kinetic}}$
where $H_{\mathrm{kinetic}}$ is the kinetic energy, $H_{\mathrm{kinetic}}=%
\frac{\Lambda }{2g}(\mathbf{\nabla n})^{2}$. A new scale $l^{2}=\frac{%
a^{2}\Lambda }{2g\kappa }$ appears ($a$ is lattice constant). Now, the core
energy is estimated as $E_{\mathrm{core}}=\mathrm{const}\cdot (\frac{\lambda
}{l})^{2}$ which indicates the instability of the static soliton against
collapse, $\lambda \rightarrow 0$. The half-skyrmion without a core $\lambda
\rightarrow 0$ turns into a spin-vortex as
\begin{equation}
\mathbf{n}_{\mathrm{hs}}=\left( \frac{x-x_{0}}{\sqrt{|\mathbf{r}-\mathbf{r}%
_{0}|^{2}+\Lambda ^{-1}}},\text{ }\pm \frac{y-y_{0}}{\sqrt{|\mathbf{r}-%
\mathbf{r}_{0}|^{2}+\Lambda ^{-1}}},\text{ }0\right) .
\end{equation}%
See Fig.4, all spins of the spin-vortex are suppressed onto the $XY$\ plane.
On the other hand, for the easy-axis anisotropic case, $\kappa <0,$ the
energy of spin-vortex will diverge, $m_{hs}\sim N\kappa \rightarrow \infty $.

\begin{figure}[tbph]
\includegraphics[width = 10cm]{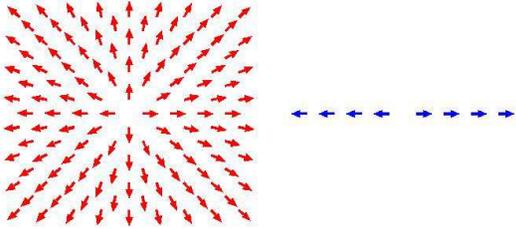}
\caption{The scheme of the spin-vortex. All spins are suppressed onto the $%
XY $\ plane. }
\end{figure}

\subsection{Fermionic zero modes}

In this part we calculate the fermionic zero modes around the spin-vortex.
By replacing the operator $\hat{c}_{i}$ and site number $i$ by Grassmann
number $\psi (x)$ and continuum coordinates $x,y$, we calculate the fermion
bound state on the spin-vortex by the continuum formula of the following
effective Hamiltonian
\begin{equation}
\mathcal{H}_{\mathrm{eff}}=-\sum\limits_{\left\langle ij\right\rangle
}(t_{i,j}\hat{c}_{i}^{\dagger }\hat{c}_{j}+h.c.)+\sum_{i}\left( -1\right)
^{i}\Delta \hat{c}_{i}^{\dagger }\mathbf{\mathbf{n}_{\mathrm{hs}}\cdot
\sigma }\hat{c}_{i}
\end{equation}%
In the continuum limit, the two-flavor Dirac-like effective Lagrangian\cite%
{st,st0,pi,wen} describes the low energy fermionic excitations at two nodes $%
\mathbf{k}_{1}=(\frac{\pi }{2},\frac{\pi }{2}),$ $\mathbf{k}_{2}=(\frac{\pi
}{2},-\frac{\pi }{2})$,
\begin{equation}
\mathcal{L}_{\mathrm{eff}}=i\bar{\psi}_{1}\gamma _{\mu }\partial _{\mu }\psi
_{1}+i\bar{\psi}_{2}\gamma _{\mu }\partial _{\mu }\psi _{2}+m_{e}(\bar{\psi}%
_{1}\mathbf{n_{\mathrm{hs}}\cdot \sigma }\psi _{1}-\bar{\psi}_{2}\mathbf{n_{%
\mathrm{hs}}\cdot \sigma }\psi _{2})
\end{equation}%
where $\bar{\psi}_{1}=\psi _{1}^{\dagger }\gamma _{0}=(%
\begin{array}{llll}
\bar{\psi}_{\uparrow 1A}, & \bar{\psi}_{\uparrow 1B}, & \bar{\psi}%
_{\downarrow 1A}, & \bar{\psi}_{\downarrow 1B}%
\end{array}%
)$ and $\bar{\psi}_{2}=\psi _{2}^{\dagger }\gamma _{0}=(%
\begin{array}{llll}
\bar{\psi}_{\uparrow 2B}, & \bar{\psi}_{\uparrow 2A}, & \bar{\psi}%
_{\downarrow 2B}, & \bar{\psi}_{\downarrow 2A}%
\end{array}%
).$ $\gamma _{\mu }$ is defined as $\gamma _{0}=\sigma _{0}\otimes \tau
_{z}, $ $\gamma _{1}=\sigma _{0}\otimes \tau _{y},$ $\gamma _{2}=\sigma
_{0}\otimes \tau _{x},$ $\sigma _{0}=\left(
\begin{array}{ll}
1 & 0 \\
0 & 1%
\end{array}%
\right) $. $\tau ^{x},$ $\tau ^{y},$ $\tau ^{z}$ are Pauli matrices. $m_{e}$
denoted the mass gap of the electrons. We have set the Fermi velocity to be
unit $v_{F}=1$. The solution of zero modes has given by%
\begin{equation*}
\psi _{1}^{0}(\mathbf{r})=\left(
\begin{array}{l}
0 \\
\exp (-\frac{\mid \mathbf{r}-\mathbf{r}_{0}\mid }{m_{e}}) \\
\exp (-\frac{\mid \mathbf{r}-\mathbf{r}_{0}\mid }{m_{e}}) \\
0%
\end{array}%
\right) \text{ and }\psi _{2}^{0}(\mathbf{r})=\left(
\begin{array}{l}
0 \\
-\exp (-\frac{\mid \mathbf{r}-\mathbf{r}_{0}\mid }{m_{e}}) \\
\exp (-\frac{\mid \mathbf{r}-\mathbf{r}_{0}\mid }{m_{e}}) \\
0%
\end{array}%
\right)
\end{equation*}%
in Ref.\cite{zeromode,kou}.

On the other hand, the double zero modes of a spin-vortex on a $21\times 21$
lattice was shown in Fig.5. The exact charge is localized around the defect
center within a length-scale $\sim \Delta ^{-1}.$ For the two spin-vortices,
the fermion zero modes are slightly split due to tunneling between them.
From numerical results, we find that there exist two zero modes on each
spin-vortex. The fermion zero modes around a $\pi $ spin-vortex match the
results in \cite{zeromode,kou}.

\begin{figure}[tbph]
\includegraphics[width = 9.0cm]{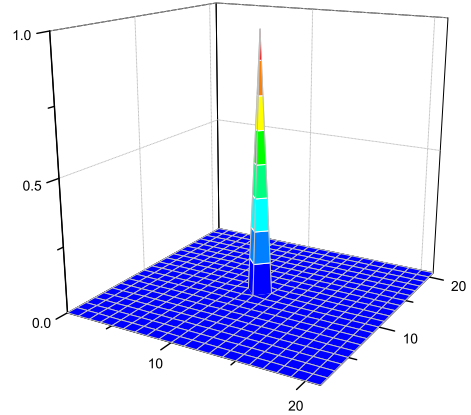}
\caption{The fermionic zero mode of spin-vortex on a $21\times 21$ lattice. }
\end{figure}

\subsection{Induced quantum numbers and statistics}

Next, we calculate the induced quantum number on the spin-vortices and show
the relationship between the induced quantum numbers and the topological
charge of them\cite{jackiw}.

For the solutions of zero modes, there are four zero-energy soliton states
around a spin-vortex which are denoted by
\begin{eqnarray}
&\mid &a_{1}\rangle \otimes \mid b_{1}\rangle ,\text{ }\mid a_{0}\rangle
\otimes \mid b_{0}\rangle , \\
&\mid &a_{0}\rangle \otimes \mid b_{1}\rangle ,\text{ }\mid a_{1}\rangle
\otimes \mid b_{0}\rangle .  \notag
\end{eqnarray}%
Here $\mid a_{0}\rangle $ and $\mid b_{0}\rangle $ are empty states of the
zero modes $\psi _{1}^{0}(\mathbf{r})$ and $\psi _{2}^{0}(\mathbf{r});$ $%
\mid a_{1}\rangle $ and $\mid b_{1}\rangle $ are occupied states of them.
Without doping, the soliton states of a spin-vortex $\mid \mathrm{sol}%
\rangle $ are denoted by $\mid a_{0}\rangle \otimes \mid b_{1}\rangle $ and $%
\mid a_{1}\rangle \otimes \mid b_{0}\rangle $.

In Ref.\cite{kou}, the induced quantum numbers on the solitons states
including total induced fermion number $\hat{N}_{F}=\sum\limits_{\alpha ,i}%
\hat{c}_{i}^{\dagger }\sigma _{z}\hat{c}_{i}$ and the induced staggered spin
number $\hat{S}_{(\pi ,\pi )}^{z}=\frac{1}{2}\sum\limits_{i\in A}\hat{c}%
_{i}^{\dagger }\sigma _{z}\hat{c}_{i}-\frac{1}{2}\sum\limits_{i\in B}\hat{c}%
_{i}^{\dagger }\sigma _{z}\hat{c}_{i}$ has been calculated. The total
induced fermion number on the solitons is zero from the cancelation effect
between two nodals $\hat{N}_{F}\mid \mathrm{sol}\rangle =0.$ However, there
exists an\ induced staggered spin moment on the soliton states\cite{kou}, $%
\hat{S}_{(\pi ,\pi )}^{z}\mid \mathrm{sol}\rangle =\pm \frac{1}{2}\mid
\mathrm{sol}\rangle .$

In particular, we will show the relationship between the induced staggered
spin moment $S_{(\pi ,\pi )}^{z}$ and the topological charge of the
spin-vortex $\mathcal{Q}$. Remember we have study the zero modes and the
induced quantum number on a spin-vortex. Now the spin polarization inside
the core of the spin-vortex will be determined by the induced spin moment :
for the case of $S_{(\pi ,\pi )}^{z}=\frac{1}{2},$ one has a up-spin
polarized core, $\mathcal{Q}\mathbf{=-}\frac{1}{2};$ for the case of $%
S_{(\pi ,\pi )}^{z}=-\frac{1}{2},$ one has a down-spin polarized core, $%
\mathcal{Q}\mathbf{=}\frac{1}{2}.$ \emph{As a result, the spin-vortex with
induced staggered spin moment becomes a half-skyrmion with a narrow core and
the topological charge of such half-skyrmion dependents on the induced spin
moment} (See Fig.6).

\begin{figure}[tbph]
\includegraphics[width = 7.0cm]{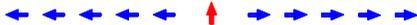}
\caption{The spin-vortex with trapped spin moment turns into a half
skyrmion. The trapped spin moment is denoted by the red arrow.}
\end{figure}

On the other hand, since each bosonic spinon\ $\mathbf{z}$ carries $\frac{1}{%
2}$ staggered spin moment, an induced staggered spin moment corresponds to a
trapped bosonic spinon $\mathbf{z}$. So there is a mutual semion statistics
between \textquotedblright bosonic spinon\textquotedblright\ $\mathbf{z}$
and the half skyrmion with narrow core (meron or antimeron ). Due to the
mutual semion statistics, by binding the trapped bosonic spinon $\mathbf{z,}$
a mobile half-skyrmion becomes a composite fermionic particle.

In order to giving a clear comparation we show these quantum numbers of a
spin vortex (half-skyrmion with narrow core) in Table.1.%
\begin{equation*}
\begin{tabular}[t]{|c|c|c|}
\hline
& $\mid a_{1}\rangle \otimes \mid b_{0}\rangle $ & $\mid a_{0}\rangle
\otimes \mid b_{1}\rangle $ \\ \hline
Total fermion number & $0$ & $0$ \\ \hline
staggered spin number & $-\frac{1}{2}$ & $\frac{1}{2}$ \\ \hline
Topological charge & $\frac{1}{2}$ & $-\frac{1}{2}$ \\ \hline
\end{tabular}%
\end{equation*}

\vskip1.5cm

In the long range AF ordered state ($g<g_{c}$), the spin-vortex will be
confined and the total energy of it diverges. On the other hand, in the
region of $g>g_{c}$, the spin-vorterx has finite energy that is $E_{\mathrm{%
core}}=\kappa $. Thus in the quantum non-magnetic insulator state ($g>g_{c}$%
), the mass of half-skyrmion vanishes in the isotropic limit\cite{au}, $%
m_{hs}=0$ and survives when there is a small easy-plane anisotropic term $%
m_{hs}=E_{\mathrm{core}}=\kappa $. For the easy-axis anisotropic case, $%
\kappa <0,$ the energy of spin-vortex always diverge, $m_{hs}\rightarrow
\infty $.

\section{Quantum spin liquid state}

Whether the non-magnetic insulator is a VBC state? Because the VBC state is
characterized by the condensation of the half-skyrmions, if half-skyrmions
are bosons, they will condense. Then one gets a VBC state (or quantum dimer
state)\cite{sac,sen}. However, in the non-magnetic insulator half-skyrmion
obeys fermionic statistics by trapping a bosonic spinon $\mathbf{z}$ onto
its core. The massless fermionic vortices lead to a new story of the quantum
spin liquid states. Consequently, a new type of quantum spin liquid state -
nodal spin liquid\emph{\ }(NSL) is explored. In addition, after adding a
small easy-plane anisotropic term, the fermionic vortices get energy gap,
then the ground state becomes a topologically ordered spin liquid.

\subsection{Effective model of half-skyrmions}

Because the spin-vortices may have zero energy, to learn the low energy
physics of non-magnetic insulator, we need to know quantum dynamics of them.
In the following parts we will consider the spin-vortex as a quantum object
and\ obtain its effective model.

As shown in Fig.6, the center of a spin-vortex is the plaquette rather than
the original sites. So we may define dual lattice by $I$. The spin-vortex
will hop from one dual lattice to another. Thus the spin-vortices show
similar behavior of vortices in XY\ model : it can move on dual lattice with
the same lattice constant and feel an effective $\pi $-flux phase\cite%
{ng,half} (See detail in Appendix C).

We may use the operator $f_{I\sigma }$ to describe such neutral fermionic
particle with half spin at dual lattice $I$. Here $\sigma $ is the spin
index. The relation between the zero energy states and the fermionic states
is given as
\begin{equation}
\mid a_{1}\rangle \otimes \mid b_{0}\rangle =f_{I\downarrow }^{\dagger }\mid
0\rangle _{f}
\end{equation}%
$\text{ and }$%
\begin{equation}
\mid a_{0}\rangle \otimes \mid b_{1}\rangle =f_{I\uparrow }^{\dagger }\mid
0\rangle _{f}
\end{equation}%
(The state $|0\rangle _{f}$ is defined through $f_{I\uparrow }|0\rangle
_{f}=f_{I\downarrow }|0\rangle _{f}=0$). We call such neutral object at dual
lattice $I$ (fermion with $\pm \frac{1}{2}$ spin degree freedom) a
"fermionic spinon". Then the leading order of the hopping term of the
half-skyrmions is
\begin{equation}
\mathcal{H}_{hs}=-\sum\limits_{\langle I,J\rangle }\left( \tilde{t}%
_{I,J}f_{I}^{\dagger }f_{J}+h.c.\right)
+\sum\limits_{I}m_{hs}(-1)^{I}f_{I}^{\dagger }f_{I}
\end{equation}%
where $f_{I}=(f_{I\uparrow },f_{I\downarrow })^{T}$ are defined as fermionic
spinon's annihilation operators. $I$ and $J$ denote two nearest-neighbor
dual sites. Here $\tilde{t}_{I,J}$ is the effective hopping of the fermionic
spinons. For the $\pi $-flux phase for the half-skyrmions, one also needs to
divide the dual-lattice into two sublattices, $A$ and $B$. After
transforming the hopping term into momentum space, we obtain a d-wave like
dispersion of the fermionic spinons $\epsilon _{\mathbf{k}}=\pm \sqrt{4%
\tilde{t}^{2}\left( \cos ^{2}p_{x}+\cos ^{2}p_{y}\right) }$. So there also
exist two nodal fermi-points at $\mathbf{p}_{1}=(\frac{\pi }{2},\frac{\pi }{2%
}),$ $\mathbf{p}_{2}=(\frac{\pi }{2},-\frac{\pi }{2})$ and the spectrum of
fermions spinons becomes linear in the vicinity of the two nodal points.

In the continuum limit, we get a Dirac-like effective Lagrangian that
describes the low energy fermionic spinons
\begin{equation}
\mathcal{L}_{hs}=\sum_{a}i\bar{\Psi}_{a}\gamma _{\mu }\partial _{\mu }\Psi
_{a}+im_{hs}\sum_{a}\overline{\Psi }_{a}\Psi _{a}
\end{equation}%
where $\bar{\Psi}_{1}=\Psi _{1}^{\dagger }\gamma _{0}=(%
\begin{array}{llll}
\bar{f}_{\uparrow 1A}, & \bar{f}_{\uparrow 1B}, & \bar{f}_{\downarrow 1A}, &
\bar{f}_{\downarrow 1B}%
\end{array}%
)$ and $\bar{\Psi}_{2}=\Psi _{2}^{\dagger }\gamma _{0}=(%
\begin{array}{llll}
\bar{f}_{\uparrow 2B}, & \bar{f}_{\uparrow 2A}, & \bar{f}_{\downarrow 2B}, &
\bar{f}_{\downarrow 2A}%
\end{array}%
).$ $\gamma _{\mu }$ is defined as $\gamma _{0}=\sigma _{0}\otimes \tau
_{z}, $ $\gamma _{1}=\sigma _{0}\otimes \tau _{y},$ $\gamma _{2}=\sigma
_{0}\otimes \tau _{x},$ $\sigma _{0}=\left(
\begin{array}{ll}
1 & 0 \\
0 & 1%
\end{array}%
\right) $. $\tau ^{x},$ $\tau ^{y},$ $\tau ^{z}$ are Pauli matrices. For
simplicity, we set the Fermi velocity to be unit. Thus in the isotropic
limit $\kappa \rightarrow 0$, we may ignore the mass gap of $\Psi _{a}$ and
consider the fermionic spinons as gapless particles.

\subsection{Mutual Chern-Simons theory}

From above results, there exist two types of fields, the bosonic spinon%
\textbf{\ }$\mathbf{z}$ and the fermionic spinon $f_{\sigma }^{\dagger }.$
Now the low energy effective model becomes%
\begin{equation}
\mathcal{L}_{\mathrm{eff}}=\mathcal{L}_{hs}+\mathcal{L}_{s}=\frac{1}{2g}%
\left( \partial _{\mu }\mathbf{n}\right) ^{2}+i\sum_{a\sigma }\overline{\Psi
}_{a\sigma }\left( \gamma _{\mu }\partial _{\mu }+m_{hs}\right) \Psi
_{a\sigma }.
\end{equation}%
However, there exists non-trivial topological relationship between $\mathbf{z%
}$ and $\Psi _{a\sigma }$ - the fields $\Psi _{a\sigma }$ that carry $\pm
\frac{1}{2}$ winding number of AF vector $\mathbf{n}$,
\begin{equation}
\frac{1}{8\pi }\epsilon _{\mu \nu \lambda }\mathbf{n}\cdot \partial _{\nu }%
\mathbf{n}\times \partial _{\lambda }\mathbf{n}=-i\sum_{a}\bar{\Psi}_{a}%
\frac{\sigma ^{z}}{2}\gamma _{\mu }\Psi _{a}.  \label{c}
\end{equation}%
Here the operator $\frac{\sigma ^{z}}{2}$ means that the topological charge
of the fermionic spinons is spin-dependence : For the meron with $\frac{1}{2}
$ ($-\frac{1}{2}$) spin number, the topological charge $\mathcal{Q}\mathbf{=-%
}\frac{1}{2}$ ($\mathcal{Q}\mathbf{=}\frac{1}{2}$) with up-spin (down-spin)
polarized in the core.

To ensure such constraint in Eq.(\ref{c}), we add a new term in the
effective Lagrangian,
\begin{equation}
\mathcal{L}_{\mathrm{constraint}}=iA_{\mu }(\frac{1}{8\pi }\epsilon _{\mu
\nu \lambda }\mathbf{n}\cdot \partial _{\nu }\mathbf{n}\times \partial
_{\lambda }\mathbf{n}+i\sum_{a}\bar{\Psi}_{a}\frac{\sigma ^{z}}{2}\gamma
_{\mu }\Psi _{a}).
\end{equation}%
Finally the low energy effective theory is obtained as
\begin{widetext}
\begin{eqnarray}
\mathcal{L}_{\mathrm{eff}} &=&\mathcal{L}_{hs}+\mathcal{L}_{s}+\mathcal{L}_{%
\mathrm{constraint}}  \notag \\
&=&i\sum_{a}\overline{\Psi }_{a}\left( \gamma _{\mu }\partial _{\mu }+i\frac{%
\sigma ^{z}}{2}\gamma _{\mu }A_{\mu }+m_{hs}\right) \Psi _{a}+\frac{1}{2g}%
\left( \partial _{\mu }\mathbf{n}\right) ^{2}+\mathcal{L}_{MCS} \\
&=&i\sum_{a}\overline{\Psi }_{a}\left( \gamma _{\mu }\partial _{\mu }+i\frac{%
\sigma ^{z}}{2}\gamma _{\mu }A_{\mu }+m_{hs}\right) \Psi _{a}+\frac{2}{g}%
|(\partial _{\mu }-ia_{\mu })\mathbf{z}|^{2}+\mathcal{L}_{MCS}.  \notag
\end{eqnarray}%
\end
{widetext} In particular, there exists a mutual-Chern-Simons term,
\begin{equation}
\mathcal{L}_{MCS}=\frac{i}{\pi }\epsilon ^{\mu \nu \lambda }A_{\mu }\partial
_{\nu }a_{\lambda }=\frac{i}{\pi }\epsilon ^{\mu \nu \lambda }a_{\mu
}\partial _{\nu }A_{\lambda }.
\end{equation}%
To obtain above low energy effective theory, we have use the equation $%
\partial _{\nu }a_{\lambda }-\partial _{\lambda }a_{\nu }=\frac{1}{2}\mathbf{%
n}\cdot \partial _{\nu }\mathbf{n}\times \partial _{\lambda }\mathbf{n}.$

Then the effective model becomes an $U(1)\times U(1)$ mutual-Chern-Simons
(MCS) gauge theory. Fermions $\Psi _{a\sigma }$ couple to an $U(1)$ gauge
field $A_{\mu }$; boson spinon $\mathbf{z}$ couples to an $U(1)$ gauge field
$a_{\mu }$. The results say that $\Psi _{a\sigma }$ act as half topological
vortices for $a_{\mu }$ and $\mathbf{z}$ act as half topological vortices of
$A_{\mu }$. This effective Lagrangian proposed in here and earlier papers
\cite{weng2,kou3} retains \emph{the full symmetries of translation, parity,
time-reversal, and global spin rotation}, in contrast to the conventional
Chern-Simons theories where the second and third symmetries are usually
broken. In the following parts of the paper, we use $U(1)\times U(1)$ MCS
theory to learn the quantum non-magnetic state near Mott transition.

\subsection{AF order}

If $g<g_{c}$, the ground state has a long range N\'{e}el order, $%
\left\langle \mathbf{z}\right\rangle =\mathbf{z}_{0}\neq 0$ or $\langle
\mathbf{n}\rangle \neq 0$. Hence the effective model turns into
\begin{eqnarray}
\mathcal{L}_{\mathrm{eff}} &\simeq &\frac{2z_{0}^{2}}{g_{_{\mathrm{eff}}}}%
\mid a_{\mu }|^{2}+\frac{i}{\pi }\epsilon ^{\mu \nu \lambda }a_{\mu
}\partial _{\nu }A_{\lambda } \\
&&+i\sum_{a}\overline{\Psi }_{a}\left( \gamma _{\mu }\partial _{\mu }+i\frac{%
\sigma ^{z}}{2}\gamma _{\mu }A_{\mu }+m_{hs}\right) \Psi _{a}  \notag
\end{eqnarray}%
where $g_{_{\mathrm{eff}}}$ is renormalized coupling constant as $g_{_{%
\mathrm{eff}}}=\frac{g\cdot g_{c}}{(g_{c}-g)}$. Here the mass term of the
gauge field $a_{\mu }$ is caused by $\mathbf{z}$ condensation. After
integrating $a_{\mu }$ we get a renormalized kinetic term for gauge field $%
A_{\mu }$,
\begin{equation}
\mathcal{L}_{\mathrm{eff}}=\frac{1}{4e_{A}^{2}}(\partial _{\mu }A_{\nu
})^{2}+\mathcal{L}_{hs}
\end{equation}%
with $\frac{1}{e_{A}^{2}}=\frac{g_{_{\mathrm{eff}}}}{z_{0}^{2}\pi ^{2}}$.
Due to the instanton effect, the gauge field $A_{\mu }$ also obtains a mass
gap and a linear confinement appears for fermions\cite{confine}. The low
energy physics is dominated only by spin waves and the effective Lagrangian
becomes $\mathcal{L}_{\mathrm{eff}}=\frac{1}{2g_{_{\mathrm{eff}}}}(\partial
_{\mu }\mathbf{n})^{2}.$

At small $T$, the AF order is believed in the so-called renormalized
classical (RC) region. In the RC region, the effective Lagrangian loses the
Lorentz invariance and becomes $\mathcal{L}_{\mathrm{eff}}\rightarrow \frac{1%
}{2}\tilde{\rho}_{s}(\mathbf{\nabla n})^{2}.$ Here $\tilde{\rho}_{s}$ is the
renormalized spin stiffness as $\tilde{\rho}_{s}=\rho _{s}(1-\frac{g_{c}}{g}%
) $. Then at low temperatures, fermionic vortices are paired by the
logarithmic-attractive interaction $V(\mathbf{r})=2\pi \tilde{\rho}_{s}\ln
\frac{\left\vert \mathbf{r}\right\vert }{a_{0}}$. With the increase of
temperature, neutral vortex-antivortex pairs like those in the $XY$ model
are thermally excited, leading to a conventional contribution to the
screening effect. By the conventional Kosterlitz-Thouless (KT) theory, there
exists a \textquotedblleft deconfining\textquotedblright\ temperature, $T_{%
\mathrm{de}}\simeq \frac{\pi \tilde{\rho}_{s}}{2}.$ Above $T_{\mathrm{de}}$,
free excited fermionic vortices exist.

\subsection{Nodal spin liquid - \textbf{isotropic case} $\protect\kappa =0$}

If $g>g_{c},$ the ground state has no AF long range order. Now $\mathbf{z}$
has a mass gap $m_{z}\mathrm{,}$ or one has the massive spin-1 quanta,
\begin{equation}
\mathcal{L}_{\mathbf{s}}=\frac{1}{2g}\left[ \left\vert (\partial _{\mu
}-ia_{\mu })\mathbf{z}\right\vert ^{2}+m_{z}^{2}\mathbf{z}^{2}\right] .
\end{equation}%
For the isotropic case $\kappa =0$,\textbf{\ }the fermionic vortex has no
energy gap. Thus the effective model becomes
\begin{eqnarray}
\mathcal{L}_{\mathrm{eff}} &=&i\sum_{a}\overline{\Psi }_{a}\left( \gamma
_{\mu }\partial _{\mu }+i\frac{\sigma ^{z}}{2}\gamma _{\mu }A_{\mu }\right)
\Psi _{a} \\
&&+\frac{1}{2g}\left[ \left\vert (\partial _{\mu }-ia_{\mu })\mathbf{z}%
\right\vert ^{2}+m_{z}^{2}\mathbf{z}^{2}\right] +\frac{i}{\pi }\epsilon
^{\mu \nu \lambda }A_{\mu }\partial _{\nu }a_{\lambda }.  \notag
\end{eqnarray}%
After integrating the massless fermions and the massive bosonic spinons, the
effective model turns into
\begin{equation}
\mathcal{L}_{\mathrm{eff}}=-\frac{1}{4e_{a}^{2}}(\partial _{\nu }a_{\mu
})^{2}-\frac{1}{4e_{A}^{2}}(\partial _{\nu }A_{\mu })^{2}+\frac{i}{\pi }%
\epsilon ^{\mu \nu \lambda }a_{\mu }\partial _{\nu }A_{\lambda }
\end{equation}%
where $e_{a}^{2}\simeq 3\pi m_{z},$ $\frac{1}{4e_{A}^{2}}=\frac{1}{4\sqrt{%
p^{2}}}$ and $p^{2}=p_{\mu }^{2}$ is the momentum. Are bosonic spinons and
fermionic vortices real quasi-particle? To answer the question, we need to
calculate the energy gap of the gauge fields furthermore.

Firstly, after integrating $a_{\mu }$ we obtain a mass term for gauge field $%
A_{\mu },$
\begin{equation}
\mathcal{L}_{\mathrm{eff}}=i\sum_{a}\overline{\Psi }_{a}\left( \gamma _{\mu
}\partial _{\mu }+i\frac{\sigma ^{z}}{2}\gamma _{\mu }A_{\mu }\right) \Psi
_{a}-\frac{1}{4e_{A}^{2}}(\partial _{\nu }A_{\mu })^{2}+\frac{e_{a}^{2}}{\pi
^{2}}A_{\mu }^{2}.
\end{equation}%
Due to exchanging the gauge field $A_{\mu }$, a short range interaction is
induced between fermions. It is obvious that the short range interaction is
irrelevant. As a result, the massless Dirac particles $\Psi _{a}$ that
couple to $A_{\mu }$ become real low energy degrees of freedom.

Secondly, after integrating $A_{\mu }$ we get the effective model of $a_{\mu
}$ as the%
\begin{equation}
\mathcal{L}_{\mathrm{eff}}=-\frac{1}{4e_{a}^{2}}(\partial _{\nu }a_{\mu
})^{2}+\frac{e_{A}^{2}}{\pi ^{2}}a_{\mu }^{2}.
\end{equation}%
This Lagrangian has the schematic form $-\frac{1}{2e_{a}^{2}}p^{2}+\frac{1}{%
\pi ^{2}}p$. That means the momentum of the gauge field $a_{\mu }$ is not
zero, $p=\frac{2e_{a}^{2}}{\pi ^{2}}\neq 0$. Then the gauge field $a_{\mu }$
has a finite energy gap ($\mathcal{L}_{\mathrm{eff}}\sim \frac{2e_{a}^{2}}{%
\pi ^{4}}a_{\mu }^{2}$) and shows roton-like behavior. As a result, the
induced interaction by exchanging the gauge field $a_{\mu }$ is irrelevant
and the bosonic spinons are real quasi-particles.

Finally, the low energy effective theory of nodal spin liquid becomes
\begin{eqnarray}
\mathcal{L}_{\mathrm{eff}} &=&i\sum_{a}\bar{\Psi}_{a}\gamma _{\mu }\partial
_{\mu }\Psi _{a}+\frac{1}{2g}\left[ (\partial _{\mu }\mathbf{z}%
)^{2}+m_{z}^{2}\mathbf{z}^{2}\right]  \label{rq} \\
&&-\frac{1}{4e_{a}^{2}}(\partial _{\nu }a_{\mu })^{2}+\frac{\sqrt{p^{2}}}{%
\pi ^{2}}a_{\mu }^{2}.  \notag
\end{eqnarray}%
There are three types of quasi-particles : \emph{two flavor gapless
fermionic spinons }$\Psi _{a\sigma }$\emph{, gapped bosonic spinons }$z$%
\emph{\ and the roton-like gauge fields }$a_{\mu }$. Therefore, from the
effective theory we conclude that NSL state is stable and can be considered
as a new type of quantum spin liquid.

\subsection{Topological spin liquid - anisotropic case $\protect\kappa >0$}

For the anisotropic case $\kappa \neq 0$,\textbf{\ }the fermionic vortex has
finite energy gap. Thus the effective model becomes
\begin{eqnarray}
\mathcal{L}_{\mathrm{eff}} &=&i\sum_{a}\overline{\Psi }_{a}\left( \gamma
_{\mu }\partial _{\mu }+i\frac{\sigma ^{z}}{2}\gamma _{\mu }A_{\mu
}+m_{hs}\right) \Psi _{a}+ \\
&&+\frac{1}{2g}\left[ \left\vert (\partial _{\mu }-ia_{\mu })\mathbf{z}%
\right\vert ^{2}+m_{z}^{2}\mathbf{z}^{2}\right] +\frac{i}{\pi }\epsilon
^{\mu \nu \lambda }A_{\mu }\partial _{\nu }a_{\lambda }.  \notag
\end{eqnarray}%
After integrating the bosonic spinons $\mathbf{z}$ and fermionic spinons $%
\Psi _{a},$ there appear the kinetic terms for gauge field $a_{\mu }$ and $%
A_{\mu },$
\begin{equation}
\mathcal{L}_{\mathrm{eff}}=-\frac{1}{4e_{A}^{2}}(\partial _{\nu }A_{\mu
})^{2}-\frac{1}{4e_{a}^{2}}(\partial _{\mu }a_{\nu })^{2}+\frac{i}{\pi }%
\epsilon ^{\mu \nu \lambda }A_{\mu }\partial _{\nu }a_{\lambda }
\end{equation}%
where $e_{a}^{2}\simeq 3\pi m_{z}$ and $e_{A}^{2}\simeq \frac{3}{4}\pi
m_{hs}.$

In particular, one may find that due to the mutual CS term, the mass term of
one gauge field can be obtained by integrating the other : by integrating $%
a_{\mu },$ the effective model of $A_{\mu }$ becomes%
\begin{equation*}
\mathcal{L}_{\mathrm{eff}}=-\frac{1}{4e_{A}^{2}}(\partial _{\nu }A_{\mu
})^{2}+\frac{e_{a}^{2}}{\pi ^{2}}A_{\mu }^{2}.
\end{equation*}%
On the other hand, by integrating $A_{\mu }$ we get the effective model of $%
a_{\mu }$ as
\begin{equation*}
\mathcal{L}_{\mathrm{eff}}=-\frac{1}{4e_{a}^{2}}(\partial _{\nu }a_{\mu
})^{2}+\frac{e_{A}^{2}}{\pi ^{2}}a_{\mu }^{2}.
\end{equation*}%
Thus the gauge fields have mass gap
\begin{equation*}
m_{a}=m_{A}=\frac{e_{A}e_{a}}{2\pi }.
\end{equation*}%
Thus we find that the mutual \textrm{U(1)}$\times $\textrm{U(1)} CS theory
describes a topological ordered spin liquid without gapless excitations. Due
to exchanging the gauge fields $A_{\mu }$ and $a_{\mu }$, a short range
interaction is induced between fermionic spinons and bosonic spinons. It is
obvious that the short range interaction is irrelevant. As a result, the
spinons that couple to $A_{\mu }$ or $a_{\mu }$ become real low energy
degrees of freedom.

To show the topological properties of the topological spin liquid, we
calculate ground state degeneracy on a torus. For periodic boundary
condition, we can expand the gauge fields as
\begin{align}
(A_{x},A_{y})& =(\frac{1}{L_{x}}\Theta _{x}+\sum_{\mathbf{k}}A_{\mathbf{k}%
}^{x}e^{i\check{x}\cdot \mathbf{k}},\frac{1}{L_{y}}\Theta _{y}+\sum_{\mathbf{%
k}}A_{\mathbf{k}}^{y}e^{i\check{x}\cdot \mathbf{k}}), \\
(a_{x},a_{y})& =(\frac{1}{L_{x}}\theta _{x}+\sum_{\mathbf{k}}a_{\mathbf{k}%
}^{x}e^{i\check{x}\cdot \mathbf{k}},\frac{1}{L_{y}}\theta _{y}+\sum_{\mathbf{%
k}}a_{\mathbf{k}}^{y}e^{i\check{x}\cdot \mathbf{k}})
\end{align}%
where $\mathbf{k=}(k_{x},k_{y})=(\frac{2\pi }{L_{x}}n_{x},$ $\frac{2\pi }{%
L_{y}}n_{y})$ where $n_{x,y}$ are integers. $(A_{\mathbf{k}}^{x},A_{\mathbf{k%
}}^{y})$ and $(a_{\mathbf{k}}^{x},a_{\mathbf{k}}^{y})$ are the gauge fields
with non-zero momentum and $\left( \Theta _{x},\Theta _{y}\right) $ and $%
\left( \theta _{x},\theta _{y}\right) $ are the zero modes with zero
momentum for the gauge fields $A_{i}$ and $a_{i}$. Because the existence of
the mass gap, the degree freedoms for gauge fields with non-zero momentum $%
\left( A_{\mathbf{k}}^{x},A_{\mathbf{k}}^{y}\right) \ $and $\left( a_{%
\mathbf{k}}^{x},a_{\mathbf{k}}^{y}\right) $ have nothing to do with the low
energy physics. The low energy physics is determined by $\left( \Theta
_{x},\Theta _{y}\right) $ and $\left( \theta _{x},\theta _{y}\right) $.

For the temporal gauge $A_{0}=0$, after the mode expansion, we write down
the following effective Hamiltonian to describe the low energy physics of
the mutual \textrm{U(1)}$\times $\textrm{U(1)} CS theory
\begin{equation}
H_{\mathrm{eff}}=\frac{(P_{\Theta _{x}}-\frac{\theta _{y}}{\pi })^{2}}{2M_{x}%
}+\frac{p_{\theta _{y}}^{2}}{2m_{y}}+\frac{(p_{\theta _{x}}-\frac{\Theta _{y}%
}{\pi })^{2}}{2m_{x}}+\frac{P_{\Theta _{y}}^{2}}{2M_{y}}  \notag
\end{equation}%
where the conjugate momentum for $\left( \Theta _{x},\Theta _{y}\right) $
and $\left( \theta _{x},\theta _{y}\right) $ are defined as%
\begin{equation*}
P_{\Theta _{x}}=M_{x}\dot{\Theta}_{x}+\frac{\theta _{y}}{2\pi },\text{ }%
P_{\Theta _{y}}=M_{y}\dot{\Theta}_{y}-\frac{\theta _{x}}{2\pi }
\end{equation*}%
and
\begin{equation*}
p_{\theta _{x}}=m_{x}\dot{\theta}_{x}+\frac{\Theta _{y}}{2\pi },\text{ }%
p_{\theta _{y}}=m_{y}\dot{\theta}_{y}-\frac{\Theta _{x}}{2\pi }.
\end{equation*}%
The masses are given as $M_{x}=\frac{1}{e_{A}^{2}}\frac{L_{y}}{L_{x}},$ $%
M_{y}=\frac{1}{e_{A}^{2}}\frac{L_{x}}{L_{y}}$ and $m_{x}=\frac{1}{e_{a}^{2}}%
\frac{L_{y}}{L_{x}},$ $m_{y}=\frac{1}{e_{a}^{2}}\frac{L_{x}}{L_{y}}.$ Thus
we map the original mutual \textrm{U(1)}$\times $\textrm{U(1)} CS theory to
a quantum mechanics model of two particles in two dimensions. The effective
Hamiltonian of $\left( \Theta _{x},\Theta _{y}\right) $ and $\left( \theta
_{x},\theta _{y}\right) $ corresponds to two particles on a torus through
two-unit flux : $(\Theta _{x},\theta _{y})$ are the coordinates of the first
particle, and $(\Theta _{y},\theta _{x})$ are the coordinates of the second
particle. The degeneracy for $(\Theta _{x},\theta _{y})$ degrees of freedom
and the degeneracy for $(\Theta _{y},\theta _{x})$ degrees of freedom are
given as $D_{(\Theta _{x},\theta _{y})}=2$ and $D_{(\Theta _{y},\theta
_{x})}=2.$ As a result, for the mutual \textrm{U(1)}$\times $\textrm{U(1)}
CS theory, the ground states have four-fold degeneracy. That means the
ground state of the anisotropic case $\kappa >0$ is a $Z_{2}$ topological
spin liquid.

For the easy-axis anisotropic case, $\kappa <0,$ the situation changes. For
this case, the effective nonlinear $\sigma $ model is not available for the
temperature below the energy scale of the anisotropic $\kappa $. Thus the
ground state are always long range (Ising) AF order in the insulator phase $%
M\neq 0.$

\subsection{Experimental predictions}

Firstly, we discuss the experimental predictions in the NSL.

In NSL, the spin-correlation decays exponentially
\begin{equation}
\left\langle S^{+}(x,y)S^{-}(0)\right\rangle =e^{i\mathbf{Q}\cdot \mathbf{R}%
_{i}}\left\langle n^{+}(x,y)n^{-}(0)\right\rangle \sim e^{i\mathbf{Q}\cdot
\mathbf{R}_{i}}e^{-r/\xi }
\end{equation}%
with $\mathbf{Q=(\pi ,\pi ),}$ $r=\sqrt{x^{2}+y^{2}}$ and $n^{\pm }=n^{x}\pm
in^{y}.$ Here $\xi $ is spin correlated length, $\xi =\frac{8\pi }{\Lambda
g_{\mathrm{eff}}}.$ In contrast, in algebraic spin liquid or algebraic
vortex liquid, the spin-correlation shows critical behavior.

Secondly, we calculate the special heat. Because the fermionic spinons have
no energy gap, the special heat is dominated by them. So at low temperature,
the special heat is
\begin{equation}
C_{V}=\frac{12\zeta }{\pi }k_{B}^{2}T^{2}
\end{equation}%
where $\zeta =\int_{0}^{\infty }\frac{x^{2}dx}{e^{x}+1}=\frac{3}{4}\Gamma
(3)\zeta (3)\simeq 1.803$.

Thirdly, we calculate the spin susceptibility. The definition of the spin
susceptibility is
\begin{equation}
F=F(B=0)-\frac{1}{2}\chi B^{2}...
\end{equation}%
where $F$ is free energy and $B$ is the external magnetic field. There are
two contributions to the total spin susceptibility $\chi $, one from bosonic
spinons, the other is from the fermionic spinons as
\begin{equation}
\chi =\chi _{\mathrm{b}}+\chi _{\mathrm{f}}.
\end{equation}%
Here $\chi _{\mathrm{b}}$ is given by\cite{sech,kou'}
\begin{equation}
\chi _{\mathrm{b}}=\frac{2}{3}\chi _{\perp }\mu _{\mathrm{B}}^{2}M^{2}+\frac{%
2\left( 2\mu _{\mathrm{B}}M\right) ^{2}}{\pi \beta }[\frac{m_{s}\beta }{%
1-e^{-m_{s}\beta }}-\ln (e^{m_{s}\beta }-1)]
\end{equation}%
with $\beta \equiv 1/k_{\mathrm{B}}T$. The contribution from the fermionic
spinons is almost linear temperature dependence as
\begin{equation}
\chi _{\mathrm{f}}=\frac{4\ln 2}{\pi \beta }
\end{equation}%
in unit of $(g\mu _{\mathrm{B}})^{2}.$

On the other hand, for TSL, the spin-correlation also decays exponentially $%
\left\langle S^{+}(x,y)S^{-}(0)\right\rangle \sim e^{i\mathbf{Q}\cdot
\mathbf{R}_{i}}e^{-r/\xi }$ with $\xi =\frac{8\pi }{\Lambda g_{\mathrm{eff}}}%
.$ However, due to mass gap, $m_{hs}\neq 0,$ at low temperature, the special
heat and the spin susceptibility from the fermionic spinons are all
proportion to $e^{-\beta m_{hs}}$ and disappear at zero temperature.

\section{Conclusion}

Let us draw a conclusion. In this paper, we study the non-magnetic insulator
state near Mott transition of 2D $\pi $-flux Hubbard model on square lattice
and find that for the isotropic case such non-magnetic insulator state is
quantum spin liquid state with nodal fermionic excitations - NSL; for the
anisotropy case it is TSL with full gapped excitations. The low energy
physics is basically determined by its $U(1)\times U(1)$ mutual Chern-Simons
gauge theory. There exist both fermionic spinons and bosonic spinons. And it
is just the mutual semion statistics between fermionic spinons and bosonic
spinons that guarantee the stability of quantum spin liquid states.

Because NSL state represents a new class of quantum state which may be
applied\ to learn the nature of the spin liquid state in other systems, for
example, the Hubbard model on honeycomb lattice. People have proposed that
quantum non-magnetic state of the Hubbard model on honeycomb lattice is
really a $Z_{2}$ topological spin liquid ordered state, which is robust
against arbitrary perturbation including the anisotropic term\cite%
{z21,z22,z23,z24,z25}. However, our results of the quantum spin liquid state
near the Mott transition of the $\pi $-flux Hubbard model are different -
for the isotropic case,\textbf{\ }the non-magnetic state with \textrm{SU(2) }%
spin rotation symmetry is nodal spin liquid (NSL) with gapless fermionic
excitations and roton-like excitations;\emph{\ }for the anisotropic case by
adding arbitrary small easy-plane anisotropic term, the non-magnetic state
becomes a $Z_{2}$ topological spin liquid with topological degenerate ground
states.\emph{\ }People may check the different theories by QMC approach in
the future.

Finally, we give a comparison on different quantum orders with $\pi $-vortex
(half-skyrmion or vison) and bosonic spinon. In general, for a system with $%
\pi $-vortex and bosonic spinon, there exist five types of quantum orders as%
\cite{xu} :

\begin{enumerate}
\item \textit{VBC state}: If one has gapless bosonic $\pi$-vortices and
massive bosonic spinons, then the ground state is always VBC state with
spontaneous translation symmetry breaking;

\item \textit{AF order}: If one has massive bosonic $\pi $-vortices and
gapless bosonic spinons, then the ground state is an SDW order with
spontaneous spin rotation symmetry breaking;

\item \textit{Algebraic vortex liquid}: if one has gapless fermionic $\pi $%
-vortices and massive bosonic spinons, then ground state may be an algebraic
vortex liquid (AVL)\cite{fisher}. In algebraic vortex liquid, fermionic
excitations themselves couple to massless $U(1)$ gauge fields, as gives an
example to algebraic spin liquid.

\item \textit{Topological spin liquid}: if one has massive fermionic $\pi $%
-vortex and massive bosonic spinon, the ground state is a topological order
with topologically degenerate ground state;

\item \textit{Nodal spin liquid}: if one has gapless fermionic $\pi $%
-vortices (the fermionic spinons in this paper) and massive bosonic spinons,
the ground state is a nodal spin liquid without any spontaneous symmetry
breaking. In particular, there exist gapped roton-like gauge modes.
\end{enumerate}

In order to giving a clear comparison, we give Table.2 :%
\begin{equation*}
\begin{tabular}[t]{|c|c|c|}
\hline
& $\pi $-vortex & bosonic spinon \\ \hline
VBC state & massless Boson & massive Boson \\ \hline
AF order & massive Boson & massless Boson \\ \hline
AVL state & massless fermion & massive Boson \\ \hline
TSL order & massive Boson(fermion) & massive Boson \\ \hline
NSL state & massless fermion & massive Boson \\ \hline
\end{tabular}%
\end{equation*}

\vskip 0.5cm

We give a short remark on the relation between NSL state and the AVL in Ref.
\cite{wen1,he,sen3}. In AVL, fermionic excitations coupling to a massless $%
U(1)$ gauge field and cannot be real quasi-particles. In NSL state, due to
the protection from the mutual semion statistics between fermionic spinons
and bosonic spinons, fermionic excitations are real excitations. There is
neither bosonic spinon nor roton-like gauge mode in AVL state and the low
energy effective theory of AVL state is also difference from that of NSL. In
AVL, the spin-correlation shows critical behavior, while \emph{in NSL state,
although there exist gapless fermionic spinons, the spin-correlation decays
exponentially}.

This research is supported by SRFDP, NFSC Grant no. 10874017 and
National Basic Research Program of China (973 Program) under the
grant No. 2011CB921803.

\section{Appendix A: Effective $\mathrm{NL}\protect\sigma \mathrm{M}$}

In this appendix we use the path-integral formulation of electrons with spin
rotation symmetry to obtain the effective $\mathrm{NL}\sigma \mathrm{M}$ of
the spin fluctuations\cite{wen,dup,dupuis,Schulz,weng,kou,kou1}. The
interaction term in Eq.[1] can be handled by using the SU(2) invariant
Hubbard-Stratonovich decomposition in the arbitrary on-site unit vector $%
\mathbf{\Omega }_{i}$
\begin{equation}
\hat{n}_{i\uparrow }\hat{n}_{i\downarrow }=\frac{\left( \hat{c}_{i}^{\dagger
}\hat{c}_{i}\right) ^{2}}{4}-\frac{1}{4}[\mathbf{\Omega }_{i}\mathbf{\cdot }%
\hat{c}_{i}^{\dag }\mathbf{\sigma }\hat{c}_{i}]^{2}.
\end{equation}%
Here $\mathbf{\sigma =}\left( \sigma _{x},\sigma _{y},\sigma _{z}\right) $
are the Pauli matrices. By replacing the electronic operators $\hat{c}%
_{i}^{\dagger }$ and $\hat{c}_{j}$ by Grassmann variables $c_{i}^{\ast }$
and $c_{j}$, the effective Lagrangian of the 2D generalized Hubbard model at
half filling is obtained:
\begin{equation}
\mathcal{L}_{\mathrm{eff}}=\sum_{i}c_{i}^{\ast }\partial _{\tau
}c_{i}-\sum\limits_{\left\langle ij\right\rangle }\left( t_{i,j}c_{i}^{\ast
}c_{j}+h.c.\right) -\Delta \sum_{i}c_{i}^{\ast }\mathbf{\Omega }_{i}\mathbf{%
\cdot \sigma }c_{i}.
\end{equation}

To describe the spin fluctuations, we use the Haldane's mapping \cite%
{dupuis,Haldane,Auerbach}:
\begin{equation}
\mathbf{\Omega}_{i}=(-1)^{i}\mathbf{n}_{i}\sqrt{1-\mathbf{L}_{i}^{2}}+%
\mathbf{L}_{i}
\end{equation}
where $\mathbf{n}_{i}=(n_{i}^{x},n_{i}^{y},n_{i}^{z})$ is the Neel vector
that corresponds to the long-wavelength part of $\mathbf{\Omega}_{i}$ with a
restriction $\mathbf{n}_{i}^{2}=1.$ $\mathbf{L}_{i}$ is the transverse
canting field that corresponds to the short-wavelength parts of $\mathbf{%
\Omega}_{i}$ with a restriction $\mathbf{L}_{i}\cdot\mathbf{n}_{i}=0$. We
then rotate $\mathbf{\Omega}_{i}$ to $\mathbf{\hat{z}}$-axis for the spin
indices of the electrons at $i$-site:\cite%
{wen,dup,dupuis,Schulz,weng,kou,kou1}%
\begin{align}
\psi_{i} & =U_{i}^{\dagger}c_{i}  \notag \\
U_{i}^{\dagger}\mathbf{n}_{i}\cdot\mathbf{\sigma}U_{i} & =\mathbf{\sigma }%
_{z}  \notag \\
U_{i}^{\dagger}\mathbf{L}_{i}\cdot\mathbf{\sigma}U_{i} & =\mathbf{l}_{i}\cdot%
\mathbf{\sigma}
\end{align}
where $U_{i}\in$\textrm{SU(2)/U(1)}.

One then can derive the following effective Lagrangian after such spin
transformation:
\begin{align}
\mathcal{L}_{\mathrm{eff}} & =\sum\limits_{i}\psi_{i}^{\ast}\partial_{\tau
}\psi_{i}+\sum\limits_{i}\psi_{i}^{\ast}a_{0}\left( i\right) \psi _{i}
\notag \\
& -\sum\limits_{<ij>}(t_{i,j}\psi_{i}^{\ast}e^{ia_{ij}}\psi_{j}+h.c.)  \notag
\\
& -\Delta\sum\limits_{i}\psi_{i}^{\ast}\left[ (-1)^{i}\mathbf{\sigma}_{z}%
\sqrt{1-\mathbf{l}_{i}^{2}}+\mathbf{l}_{i}\cdot\mathbf{\sigma}\right]
\psi_{i}  \label{lag}
\end{align}
where the auxiliary gauge fields $a_{ij}=a_{ij,1}\sigma_{x}+a_{ij,2}\sigma
_{y}$ and $a_{0}\left( i\right) =a_{0,1}\left( i\right)
\sigma_{x}+a_{0,2}\left( i\right) \sigma_{y}\ $are defined as
\begin{equation}
e^{ia_{ij}}=U_{i}^{\dagger}U_{j},\text{ }a_{0}\left( i\right)
=U_{i}^{\dagger}\partial_{\tau}U_{i}.  \label{ao}
\end{equation}

In terms of the mean field result $M=\left( -1\right) ^{i}\langle \psi
_{i}^{\ast }\mathbf{\sigma }_{z}\psi _{i}\rangle $ as well as the
approximations,
\begin{equation}
\sqrt{1-\mathbf{l}_{i}^{2}}\simeq 1-\frac{\mathbf{l}_{i}^{2}}{2},\text{ }%
e^{ia_{ij}}\simeq 1+ia_{ij},  \notag
\end{equation}%
we obtain the effective Hamiltonian as:
\begin{align}
\mathcal{L}_{\mathrm{eff}}& \simeq \sum\limits_{i}\psi _{i}^{\ast }\partial
_{\tau }\psi _{i}+\sum\limits_{i}\psi _{i}^{\ast }[a_{0}\left( i\right)
-\Delta \mathbf{\sigma \cdot l}_{i}]\psi _{i}  \notag \\
& -\sum\limits_{\left\langle ij\right\rangle }[t_{i,j}\psi _{i}^{\ast
}(1+ia_{ij})\psi _{j}+h.c.]  \notag \\
& -\Delta \sum\limits_{i}(-1)^{i}\psi _{i}^{\ast }\sigma _{z}\psi
_{i}+\Delta M\sum\limits_{i}\frac{\mathbf{l}_{i}^{2}}{2}.
\end{align}%
By integrating out the fermion fields $\psi _{i}^{\ast }$ and $\psi _{i},$
the effective action with the quadric terms of $[a_{0}\left( i\right)
-\Delta \mathbf{\sigma \cdot l}_{i}]$ and $a_{ij}$ becomes
\begin{equation}
\mathcal{S}_{\mathrm{eff}}=\frac{1}{2}\int_{0}^{\beta }d\tau
\sum_{i}[-4\varsigma (a_{0}\left( i\right) -\Delta \mathbf{\sigma \cdot l}%
_{i})^{2}+4\rho _{s}a_{ij}^{2}+\frac{2\Delta ^{2}}{U}\mathbf{l}_{i}^{2}].
\label{efff}
\end{equation}

To give $\rho _{s}$ and $\varsigma $ for calculation, we choose $U_{i}$ to be%
\begin{equation}
U_{i}=\left(
\begin{array}{cc}
z_{i\uparrow }^{\ast } & z_{i\downarrow }^{\ast } \\
-z_{i\downarrow } & z_{i\uparrow }%
\end{array}%
\right) ,
\end{equation}%
where $\mathbf{n}_{i}=\mathbf{\bar{z}}_{i}\mathbf{\sigma z}_{i},$ $\mathbf{z}%
_{i}=\left( z_{i\uparrow },z_{i\downarrow }\right) ^{T},$ $\mathbf{\bar{z}}%
_{i}\mathbf{z}_{i}\mathbf{=1.}$ And the spin fluctuations around $\mathbf{n}%
_{i}=\mathbf{\hat{z}}_{i}$ are
\begin{align}
\mathbf{n}_{i}& =\mathbf{\hat{z}}_{i}\mathbf{+}\text{Re}\left( \mathbf{\phi }%
_{i}\right) \mathbf{\hat{x}+}\text{Im}\left( \mathbf{\phi }_{i}\right)
\mathbf{\hat{y}} \\
\mathbf{z}_{i}& =\left(
\begin{array}{c}
1-\left\vert \mathbf{\phi }_{i}\right\vert ^{2}/8 \\
\mathbf{\phi }_{i}/2%
\end{array}%
\right) +O\left( \mathbf{\phi }_{i}^{3}\right) .
\end{align}%
Then the quantities $U_{i}^{\dagger }U_{j}$ and $U_{i}^{\dagger }\partial
_{\tau }U_{i}$ can be expanded in the power of $\mathbf{\phi }_{i}-\mathbf{%
\phi }_{j}$ and $\partial _{\tau }\mathbf{\phi }_{i},$%
\begin{align}
U_{i}^{\dagger }U_{j}& =e^{-i\frac{\mathbf{\phi }_{i}-\mathbf{\phi }_{j}}{2}%
\sigma _{y}} \\
U_{i}^{\dagger }\partial _{\tau }U_{i}& =\left(
\begin{array}{cc}
0 & \frac{1}{2}\partial _{\tau }\mathbf{\phi }_{i} \\
-\frac{1}{2}\partial _{\tau }\mathbf{\phi }_{i} & 0%
\end{array}%
\right) .
\end{align}%
According to Eq.(\ref{ao}), the gauge field $a_{ij}$ and $a_{0}\left(
i\right) $ are given as
\begin{align}
a_{ij}& =-\frac{1}{2}\left( \mathbf{\phi }_{i}-\mathbf{\phi }_{j}\right)
\mathbf{\sigma }_{y} \\
a_{0}\left( i\right) & =\frac{i}{2}\partial _{\tau }\mathbf{\phi }_{i}%
\mathbf{\sigma }_{y}.
\end{align}%
Supposing $a_{ij}$ and $a_{0}\left( i\right) $ to be a constant in space and
denoting $\mathbf{\partial }_{i}\mathbf{\mathbf{\phi }}_{i}\mathbf{=a}$ and $%
\partial _{\tau }\mathbf{\phi }_{i}=iB_{y}$, we have
\begin{align}
a_{ij}& =-\frac{1}{2}\mathbf{a\cdot (i-j)\sigma }_{y} \\
a_{0}\left( i\right) & =-\frac{1}{2}B_{y}\mathbf{\sigma }_{y}.
\end{align}%
The energy of Hamiltonian of Eq.(\ref{efff}) becomes
\begin{equation}
E\left( B_{y},\mathbf{a}\right) =\frac{1}{2}\zeta B_{y}^{2}+\frac{1}{2}\rho
_{s}\mathbf{a}^{2}.  \label{ener}
\end{equation}%
Then one could get $\zeta $ and $\rho _{s}$ from the following equations by
calculating the partial derivative of the energy
\begin{align}
\zeta & =\frac{1}{N}\frac{\partial ^{2}E_{0}\left( B_{y}\right) }{\partial
B_{y}^{2}}|_{B_{y}=0} \\
\rho _{s}& =\frac{1}{N}\frac{\partial ^{2}E_{0}\left( \mathbf{a}\right) }{%
\partial \mathbf{a}^{2}}|_{\mathbf{a}=0}.  \label{wero}
\end{align}%
Here $E_{0}\left( B_{y}\right) $ and $E_{0}\left( \mathbf{a}\right) $ are
the energy of the lower Hubbard band%
\begin{align}
E_{0}\left( B_{y}\right) & =\sum\limits_{\mathbf{k}}\left( E_{+,\mathbf{k}%
}^{\zeta }+E_{-,\mathbf{k}}^{\zeta }\right)  \label{ek} \\
E_{0}\left( \mathbf{a}\right) & =\sum\limits_{\mathbf{k}}\left( E_{+,\mathbf{%
k}}^{\rho }+E_{-,\mathbf{k}}^{\rho }\right)  \label{er}
\end{align}%
where $E_{+,\mathbf{k}}^{\zeta },$ $E_{-,\mathbf{k}}^{\zeta }$ and $E_{+,%
\mathbf{k}}^{\rho },$ $E_{-,\mathbf{k}}^{\rho }$ are the energies of the
following Hamiltonian $\mathcal{H}^{\zeta }$ and $\mathcal{H}^{\rho }$%
\begin{align}
\mathcal{H}^{\zeta }& =-\sum\limits_{<ij>}\left( t_{i,j}\psi _{i}^{\ast
}\psi _{j}+h.c.\right) -\Delta \sum\limits_{i}(-1)^{i}\psi _{i}^{\ast }%
\mathbf{\sigma }_{z}\psi _{i}  \notag \\
& +\sum\limits_{i}\psi _{i}^{\ast }a_{0}\left( i\right) \psi _{i}
\label{hca}
\end{align}%
\begin{equation}
\mathcal{H}^{\rho }=-\sum\limits_{<ij>}\left( t_{i,j}\psi _{i}^{\ast
}e^{a_{ij}}\psi _{j}+h.c.\right) -\Delta \sum\limits_{i}(-1)^{i}\psi
_{i}^{\ast }\mathbf{\sigma }_{z}\psi _{i}.  \label{hrou}
\end{equation}%
Using the Fourier transformations for $\mathcal{H}^{\zeta }$ , we have the
spectrum of the lower band of $\mathcal{H}^{\zeta }$%
\begin{equation}
E_{\pm ,\mathbf{k}}^{\zeta }=-\sqrt{\left( \left\vert \xi _{\mathbf{k}%
}\right\vert \pm \frac{B_{y}}{2}\right) ^{2}+\Delta ^{2}}
\end{equation}%
where $\xi _{\mathbf{k}}=\pm \sqrt{4t^{2}\left( \cos ^{2}k_{x}+\cos
^{2}k_{y}\right) }$. And $\zeta $ is obtained as
\begin{equation}
\zeta =\frac{1}{N}\sum\limits_{\mathbf{k}}\frac{\Delta ^{2}}{4\left(
\left\vert \xi _{\mathbf{k}}\right\vert ^{2}+\Delta ^{2}\right) ^{\frac{3}{2}%
}}.  \label{pic}
\end{equation}%
Similarly, using the Fourier transformations for $\mathcal{H}^{\rho },$ we
obtain the spectrum of the lower band of $\mathcal{H}^{\rho }$
\begin{equation}
E_{\pm ,\mathbf{k}}^{\rho }=-\sqrt{\Delta ^{2}+\left\vert \vartheta
\right\vert ^{2}+\left\vert \varphi \right\vert ^{2}\pm \left[ 4\Delta
^{2}\left\vert \vartheta \right\vert ^{2}-\left( \varphi \vartheta ^{\ast
}-\vartheta \varphi ^{\ast }\right) ^{2}\right] ^{\frac{1}{2}}}
\end{equation}%
where $\varphi $ and $\vartheta $ are defined as
\begin{align}
\varphi & =-t\sum\limits_{\mathbf{\delta }}e^{i\mathbf{k\cdot \delta }}\cos
\left( \frac{1}{2}\mathbf{a\cdot \delta }\right) \\
\vartheta & =-t\sum\limits_{\mathbf{\delta }}e^{i\mathbf{k\cdot \delta }%
}\sin \left( \frac{1}{2}\mathbf{a\cdot \delta }\right)
\end{align}%
where $\mathbf{\delta =}a\mathbf{(}e_{x},$ $e_{y}\mathbf{)}$ and $%
e_{x}^{2}=e_{y}^{2}=1$. Using Eq.(\ref{wero}), $\rho _{s}$ is given as
\begin{equation}
\rho _{s}=\frac{1}{N}\sum\limits_{\mathbf{k}}\frac{\epsilon ^{2}}{%
4(\left\vert \xi _{\mathbf{k}}\right\vert ^{2}+\Delta ^{2})^{\frac{3}{2}}}.
\label{ita}
\end{equation}%
For the $\pi $-flux Hubbard model, we get the corresponding coefficient $%
\epsilon ^{2}$ as
\begin{align}
\epsilon ^{2}& =t^{2}[\cos \left( 2k_{x}\right) \left( \Delta
^{2}+8t^{2}+4t^{2}\cos \left( 2k_{y}\right) \right)  \notag \\
& +\Delta ^{2}+3t^{2}+t^{2}\cos \left( 4k_{x}\right) ].
\end{align}

Next, to learn the properties of the low energy physics, we study the
continuum theory of the effective action in Eq.(\ref{efff}). In the
continuum limit, we denote $\mathbf{n}_{i}$, $\mathbf{l}_{i}$, $%
ia_{ij}\simeq U_{i}^{\dagger }U_{j}-1$ and $a_{0}\left( i\right)
=U_{i}^{\dagger }\partial _{\tau }U_{i}$ by $\mathbf{n}(x,y)$, $\mathbf{l}%
(x,y)$, $U^{\dagger }\partial _{x}U$ (or $U^{\dagger }\partial _{y}U$) and $%
U^{\dagger }\partial _{\tau }U,$ respectively. From the relations between $%
U^{\dagger }\partial _{\mu }U$ and $\partial _{\mu }\mathbf{n,}$
\begin{align}
a_{\tau }^{2}& =a_{\tau ,1}^{2}+a_{\tau ,2}^{2}=-\frac{1}{4}(\partial _{\tau
}\mathbf{n})^{2},\text{ }\tau =0,  \label{tao} \\
a_{\mu }^{2}& =a_{\mu ,1}^{2}+a_{\mu ,2}^{2}=\frac{1}{4}(\partial _{\mu }%
\mathbf{n})^{2},\text{ }\mu =x,y,  \label{miu} \\
\mathbf{a}_{0}\mathbf{\cdot l}& \mathbf{=}\mathbf{-}\frac{i}{2}\left(
\mathbf{n}\times \partial _{\tau }\mathbf{n}\right) \cdot \mathbf{l,}
\label{dot}
\end{align}%
the continuum formulation of the action in Eq.(\ref{efff}) turns into
\begin{align}
\mathcal{S}_{\mathrm{eff}}& =\frac{1}{2}\int_{0}^{\beta }d\tau \int d^{2}%
\mathbf{r}[\varsigma (\partial _{\tau }\mathbf{n)}^{2}+\rho _{s}\left(
\mathbf{\bigtriangledown n}\right) ^{2}  \notag \\
& -4i\Delta \varsigma \left( \mathbf{n}\times \partial _{\tau }\mathbf{n}%
\right) \cdot \mathbf{l}+(\frac{2\Delta ^{2}}{U}-4\Delta ^{2}\varsigma )%
\mathbf{l}^{2}]  \label{sigma}
\end{align}%
where the vector $\mathbf{a}_{0}$ is defined as $\mathbf{a}_{0}=\left(
a_{0,1},\text{ }a_{0,2},\text{ }0\right) .$

Finally we integrate the transverse canting field $\mathbf{l}$ and obtain
the effective $\mathrm{NL}\sigma \mathrm{M}$ as
\begin{equation}
\mathcal{S}_{\mathrm{eff}}=\frac{1}{2g}\int_{0}^{\beta }d\tau \int d^{2}r[%
\frac{1}{c}\left( \partial _{\tau }\mathbf{n}\right) ^{2}+c\left( \mathbf{%
\bigtriangledown n}\right) ^{2}]\text{ }  \label{non}
\end{equation}%
with a constraint $\mathbf{n}^{2}=1.$ The coupling constant $g$ and spin
wave velocity $c$ are defined as\cite{wen,dup,dupuis,Schulz,weng,kou,kou1}:
\begin{equation}
g=\sqrt{\frac{1}{\chi ^{\perp }\rho _{s}}},\text{ \ }c^{2}=\frac{\rho _{s}}{%
\chi ^{\perp }}
\end{equation}%
where $\chi ^{\perp }\ $is the transverse spin susceptibility%
\begin{equation}
\chi ^{\perp }=[(\frac{1}{N}\sum\limits_{\mathbf{k}}\frac{\Delta ^{2}}{%
4(\left\vert \xi _{\mathbf{k}}\right\vert ^{2}+\Delta ^{2})^{\frac{3}{2}}}%
)^{-1}-2U]^{-1}.
\end{equation}

Then we use the effective $\mathrm{NL}\sigma \mathrm{M}$ to study the
magnetic properties of the insulator state. The Lagrangian\ of $\mathrm{NL}%
\sigma \mathrm{M}$ with a constraint ($\mathbf{n}^{2}=1$) by a Lagrange
multiplier $\lambda $ becomes
\begin{equation}
\mathcal{L}_{\mathrm{eff}}=\frac{1}{2cg}\left[ \left( \partial _{\tau }%
\mathbf{n}\right) ^{2}+c^{2}\left( \mathbf{\bigtriangledown n}\right)
^{2}+i\lambda (1-\mathbf{n}^{2})\right]
\end{equation}%
where $i\lambda =m_{s}^{2}$ and $m_{s}$ is the mass gap of the spin
fluctuations. Using the large-\emph{N} approximation we rescale the field $%
\mathbf{n\rightarrow }\sqrt{N}\mathbf{n}$ and obtain the saddle-point
equation of motion as \cite{cha,chu,sech}%
\begin{equation}
\left( n_{0}\right) ^{2}+k_{\mathrm{B}}T\sum_{\omega _{n},\mathbf{q\neq 0}}%
\frac{gc}{\omega _{n}^{2}+c^{2}\mathbf{q}^{2}+m_{s}^{2}}=1.  \label{eta}
\end{equation}%
In Eq.(\ref{eta}), $n_{0}$ is the mean field value of $\mathbf{n}$ and $%
\omega _{n}=2\pi nk_{\mathrm{B}}T$, $n=$ integers.

From Eq.(\ref{eta}), we may get the solution of $m_{s}$ as
\begin{equation}
m_{s}=2k_{\mathrm{B}}T\sinh ^{-1}\left[ e^{-\frac{2\pi c}{gk_{\mathrm{B}}T}%
}\sinh \left( \frac{c\Lambda }{2k_{\mathrm{B}}T}\right) \right] .
\label{mas}
\end{equation}%
At zero temperature the solutions of $n_{0}$ and $m_{s}$ of Eq.(\ref{eta})
are determined by the coupling constant $g.$ There exists a critical point $%
g_{c}=\frac{4\pi }{\Lambda }$ : For the case of $g<\frac{4\pi }{\Lambda },$
we get a non-magnetic insulator with solutions of $n_{0}$ and $m_{s}$:%
\begin{equation}
n_{0}=(1-\frac{g}{g_{c}})^{1/2}\text{, }m_{s}=0
\end{equation}%
For the case of $g>\frac{4\pi }{\Lambda },$ we get a long range AF order
with solutions of $n_{0}$ and $m_{s}$:%
\begin{equation}
n_{0}=0\text{, }m_{s}=4\pi c(\frac{1}{g_{c}}-\frac{1}{g})
\end{equation}

\section{Appendix B : The induced kinetic term of gauge field}

Starting from the \textrm{CP(1)} model,%
\begin{equation*}
\mathcal{L}_{\mathbf{s}}=\frac{1}{2g}\left[ \left\vert (\partial _{\mu
}-ia_{\mu })\mathbf{z}\right\vert ^{2}+m_{z}^{2}\mathbf{z}^{2}\right] ,
\end{equation*}%
we calculate the induced the kinetic term of the gauge field $a_{\mu }$.

We obtain the expression of the expansion of the renormalization 2-point
function at one-loop order :
\begin{equation}
\frac{2}{\left( 2\pi \right) ^{3}}\int d^{3}k\left[ {\frac{(k_{\mu }+2p_{\mu
})(k_{\nu }+2p_{\nu })}{(k^{2}+m_{z}^{2})((k+p)^{2}+m_{z}^{2})}-}2\delta
_{\mu \nu }{\frac{1}{k^{2}+m_{z}^{2}}}\right]
\end{equation}%
At low energy limit as $p^{2}\rightarrow 0,$ we simplify the above integral
to the following one
\begin{eqnarray}
&&\frac{2}{\left( 2\pi \right) ^{3}}\pi ^{1/2}\Gamma \left( \frac{1}{2}%
\right) \frac{1}{3m_{z}^{2}}\left( p_{\mu }p_{\nu }-\delta _{\mu \nu
}p^{2}\right) \\
&=&\frac{2}{24\pi ^{2}m_{z}^{2}}\left( p_{\mu }p_{\nu }-\delta _{\mu \nu
}p^{2}\right) .  \notag
\end{eqnarray}
\ At low energy limit ( $p^{2}\rightarrow 0$). Thus from $-\frac{1}{%
4e_{a}^{2}}(\partial _{\nu }a_{\mu })^{2}$, the induced coupling constants
of three dimensional gauge field is obtained as
\begin{equation}
e_{a}^{2}=3\pi m_{z}^{2}\text{.}
\end{equation}

Using similar approach, oen may get the induced kinetic term of gauge field $%
A_{\mu }$.

\section{Appendix C : Projective symmetry group of half skyrmions}

In the insulator state (non-magnetic or magnetic order), there exists a
bosonic spinon on each site. Because, the half skyrmion is really the
vortex, we may consider the dynamics of half skyrmions as that of the
vortices of the insulator state of bosons on lattice with unit filling (See
Fig.7). As a result, the square lattice of bosonic spinons plays a role of $%
\pi $-flux phase of half skyrmion on dual lattice $\mathbf{I}=(I_{x},I_{y})$%
. So we may use the approach of the projective representation of the space
group (PSG) for vortex to learn the dynamics of half skyrmion here. The PSG\
denotes the phase transformations associating the operations in the space
group.

\begin{figure}[ptb]
\includegraphics[width=0.5\textwidth]{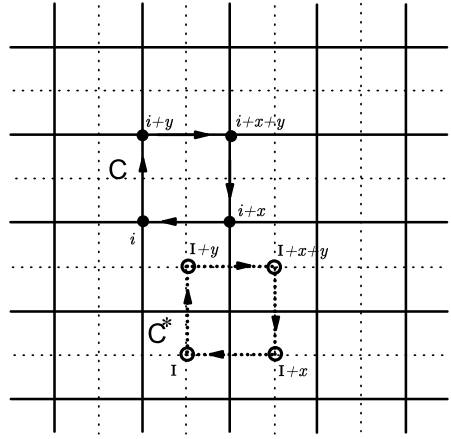}
\caption{Spin-vortex on dual lattice}
\end{figure}

For the $\pi $-flux phase of half skyrmions, we have $x$ and $y$
translations, and a $\pi /2$ rotation of the PSG on the dual square lattice
as
\begin{eqnarray}
T_{y} &:&f(I_{x},\text{ }I_{y})\rightarrow f(I_{x},\text{ }I_{y}-1)  \notag
\label{eq:psg1} \\
T_{x} &:&f(I_{x},\text{ }I_{y})\rightarrow f(I_{x}+1,\text{ }%
I_{y})(-1)^{I_{y}}  \notag \\
R_{\pi /2} &:&f(I_{x},\text{ }I_{y})\rightarrow
f(I_{x},-I_{y})(-1)^{I_{x}+I_{y}}
\end{eqnarray}%
where $f(I_{x},I_{y})$ is the wave-function of half skyrmions and $\mathbf{I}%
=(I_{x},I_{y})$ are the dual lattices. One may check that the operations
associated with translations in the PSG do not commute as $%
T_{x}T_{y}=-T_{y}T_{x}\ $that means the half skyrmion obtains a $\pi $ phase
encircling a site of the direct lattice ($i$) with one bosonic spinon.

We are looking at the half skyrmions hopping around dual square lattice in
the presence of $\pi $-flux per plaquette. Then the effective action should
be invariant under PSG and the dual $U(1)$ gauge symmetry. For the simplest
gauge and nearest neighbor hopping, the effective Hamiltonian of half
skyrmions is
\begin{equation}
\mathcal{H}_{hs}=-\sum\limits_{\langle I,J\rangle }\left( \tilde{t}_{\mathbf{%
I},\mathbf{J}}f_{\mathbf{I}}^{\dagger }f_{\mathbf{J}}+h.c.\right)
\end{equation}%
where the effective nearest-neighbor hopping $\tilde{t}_{\mathbf{I},\mathbf{J%
}}$ could be chosen as $\tilde{t}_{\mathbf{i},\mathbf{i}+\hat{x}}=t,$ $%
\tilde{t}_{\mathbf{i},\mathbf{i}+\hat{y}}=\tilde{t}e^{\pm i\frac{\pi }{2}}$.


\begin{thebibliography}{99}
\bibitem{flux} The $\pi $-flux Hubbard model (or the Hubbard model with $%
\phi $ flux) is not a physics model in condensed matter physics. However,
people pointed out that the $\pi $-flux Hubbard model may be designed with
ultracold atoms in an optical lattice. An artificial magnetic field of $\pi $%
-flux (or $\phi $ flux) in an optical square lattice is proposed to be
realized by different approaches recently\cite{2,1,3}.
\bibitem{theta} The effective \textrm{O(3)} nonlinear $\sigma $ model is
derived by integrating the fermionic field. We find that there is induced
Wess-Zumino term here. So the effective \textrm{O(3)} nonlinear $\sigma $
model don't have a Berry term as that appears in the effective model from
the Heisenberg model.
\end{thebibliography}
\end{document}